\documentclass[10pt,journal,compsoc]{IEEEtran}
\usepackage{booktabs} % For formal tables

\usepackage[ruled]{algorithm2e} % For algorithms

\SetAlFnt{\small}
\SetAlCapFnt{\small}
\SetAlCapNameFnt{\small}
\SetAlCapHSkip{0pt}
\IncMargin{-\parindent}

\usepackage{amsmath}
\usepackage{algorithmic}
\usepackage{subfigure}
\usepackage[pdftex]{graphicx}

\ifCLASSOPTIONcompsoc
  \usepackage[nocompress]{cite}
\else
  \usepackage{cite}
\fi
\ifCLASSINFOpdf

\else

\fi

\begin{document}
\title{Social-Network-Assisted Worker Recruitment in 
Mobile Crowd Sensing}

\author{Jiangtao Wang, Feng Wang, Yasha Wang, Daqing Zhang, Leye Wang, Zhaopeng Qiu

	%and~Jane~Doe,% <-this % stops a space

	\IEEEcompsocitemizethanks{\IEEEcompsocthanksitem Jiangtao Wang, Feng Wang, Yasha Wang, Daqing Zhang, and Zhaopeng Qiu are with
 School of EECS, Peking University, Beijing;\protect\\
	E-mail: jiangtaowang@pku.edu.cn;

	\IEEEcompsocthanksitem Leye Wang is with Hong Kong University of Science and Technology.

		%\IEEEcompsocthanksitem Chao Chen is with College of Computer Science, Chongqing University, China.

		%\IEEEcompsocthanksitem Qin Lv is with University of Colorado Boulder.

	}% <-this % stops an unwanted space

	\thanks{Manuscript received Sept. 30, 2017.}}


\markboth{Journal of \LaTeX\ Class Files,~Vol.~14, No.~8, August~2015}%
{Shell \MakeLowercase{\textit{et al.}}: Bare Demo of IEEEtran.cls for Computer Society Journals}



\IEEEtitleabstractindextext{%
\begin{abstract}
Worker recruitment is a crucial research problem in \textit{Mobile Crowd Sensing} (MCS). While previous studies rely on a specified platform with a pre-assumed large user pool, this paper leverages the influence propagation on the social network to assist the MCS worker recruitment. We first select a subset of users on the social network as initial seeds and push MCS tasks to them. Then, influenced users who accept tasks are recruited as workers, and the ultimate goal is to maximize the coverage. Specifically, to select a near-optimal set of seeds, we propose two algorithms, named \textit{Basic-Selector} and \textit{Fast-Selector}, respectively. \textit{Basic-Selector} adopts an iterative greedy process based on the predicted mobility, which has good performance but suffers from inefficiency concerns. To accelerate the selection, \textit{Fast-Selector} is proposed, which is based on the interdependency of geographical positions among friends. Empirical studies on two real-world datasets verify that \textit{Fast-Selector} achieves higher coverage than baseline methods under various settings, meanwhile, it is much more efficient than \textit{Basic-Selector} while only sacrificing a slight fraction of the coverage. 
\end{abstract}

}

\maketitle

\IEEEdisplaynontitleabstractindextext

% \fi
\IEEEpeerreviewmaketitle

\IEEEraisesectionheading{\section{Introduction}\label{sec:introduction}}
\IEEEPARstart{T}{he} idea of crowdsourcing rapidly mobilizes large numbers of people to work collectively for accomplishing complicated tasks \cite{1}. Recently, with the proliferation of sensor-rich mobile devices, a special form of crowdsourcing called \textit{Mobile Crowd Sensing}(MCS) \cite{3} (a.k.a., \textit{mobile crowdsourcing or participatory sensing}) has become an emerging paradigm to collaboratively collect sensing data and extract knowledge in smart cities. Different from the general crowdsourcing, MCS requires mobile users (called \textit{workers}) physically move to certain locations and collect local sensing data. By aggregating data from a group of dynamically moving workers in different subareas, the task \textit{organizers} could easily obtain a "big picture" of the sensing phenomenon (e.g., air quality, temperature or noise level) in the entire sensing target area.

The success of MCS requires the participation from a large number of workers. Thus, how to recruit a large group of workers and appropriately assign tasks is a major research problem \cite{5,10,19,13}. In particular, sensing quality (e.g., coverage) and cost (e.g. incentive budget) are two significant but opposing concerns for the organizers.  Accordingly, state-of-the-art studies propose optimal worker selection strategies to achieve a good tradeoff between sensing quality and cost. They commonly assume that workers are recruited from a pre-built and task-specific MCS system, where there is already a large pool of candidate workers. Thus, their goal is to select an appropriate subset of users from the pool with the consideration of sensing quality and cost.

However, such assumptions may not hold in real-world practices. For an MCS application, it takes the time to accumulate an adequately large user group \cite{12}. Especially when an MCS application has just been released to the public, mobile users may not even know about it. Under such circumstances, the assumed large user pool no longer exists. Thus, the above worker selection approaches fail to achieve high data quality for the MCS tasks, even if all candidate users are selected. Therefore, it is challenging to recruit workers under such a cold-start situation, which, to the best of our knowledge, has not been addressed in the previous literature for MCS.

The popularity of mobile social networks (MSN) (e.g., Facebook, Twitter, and Foursquare, etc.) has created new mediums for information sharing and propagation. Pilot studies on real-world datasets \cite{9,25,26,29} prove that it is feasible to advertise novel products or innovative ideas through the influence propagation on social networks. 

Inspired by this, instead of relying on specific MCS platforms, this paper attempts to recruit workers of MCS task in a novel manner, that is, exploiting social network as the recruitment platform. First, we select a subset of users on the MSN as initial seeds and push tasks to them. Then, influenced users accept the tasks and propagate to their friends through the spread of influence \cite{26}, and get a certain amount of incentive reward in return. Specifically, the objective of seeds selection in this paper is to maximize the temporal-spatial coverage of the MCS task obtained by finally influenced users with the following two constraints.

\begin{itemize}
	\item \textit{The number of initial seeds.} Just as advertising any other products or ideas on social networks, pushing MCS tasks to mobile users bring them intrusiveness \cite{16}. Thus, in this paper, instead of broadcasting an MCS task to a large number of users, we assume that the number of initial seeds is limited.
	\item \textit{The number of recruited workers.} For a certain MCS task, there is a total budget constraint for recruiting the workers. We assume that each influenced user would be recruited as workers and given an equal reward, thus the number of recruited users is limited. In other words, through the spread of influence on the social network, the number of finally influenced users is limited.
\end{itemize}

To further demonstrate the research issues and challenges, a motivating case is illustrated as follows. \textit{The city government has developed an MCS application, named AirSense, through which citizens can sense and report real-time AQI (Air Quality Index) in the city through the embedded sensor in their smartphones. The entire sensing area of AirSense is divided into 100 virtual subareas (e.g., 5km$\times$5km per subarea), while the entire duration is divided into several equal-length cycles (e.g., one hour per cycle). A subarea in a cycle is considered to be covered when at least one AQI reading is obtained. The total budget for worker recruitment is 10000 US dollars. As AirSense is beneficial for the public health, the social network owner is quite willing to assist its worker recruitment. Thus, we need to develop an algorithm to decide a limited number of users (e.g. at most 100) as initial seeds and push the task/application to them. Then, AirSense would be propagated through the social network with the spread of influence. Users can accept the task, download the app to report sensing data, and meanwhile recommend it to their friends. Each of these users can get a fixed equal reward (e.g. 5 US dollars) in return.}

From the above use case, we can see that the key problem is selecting a subset of users on the social network as seeds, with the objective of maximizing the temporal-spatial coverage obtained by finally influenced users under two constraints: a) the number of initial seeds should be no more than 100; b) the number of recruited workers should be no more than 10000/5=2000. Given a certain subset of users (i.e., initial seeds), we assume that the propagation of MCS task will be terminated, either when there are no new users accept the task or when the number of recruited workers reach to the limitation.

To some extent, the above-defined problem is similar to the influence maximization \cite{21,22,25,26} in social network research community, which aims at selecting a pre-defined number of seeds with the goal of maximizing the number of influenced users. However, the maximized number of collaborative MCS workers does not necessarily result in an optimal coverage. This is because users commonly spread their influence to those with similar routine trajectories \cite{15}, thus the collected data is redundant in some subareas while inadequate in others. Therefore, existing seeds selection algorithms for influence maximization cannot be directly adopted in our defined problem.

With the above mentioned research objective and challenges, the main contributions of this paper are:
\begin{itemize}
	\item Different from previous research work relying on MCS-specialized platforms with isolated users, we study a novel MCS worker recruitment problem by leveraging the spread of influence among connected users on social networks. With sensing coverage, users' intrusiveness and total incentive budget in mind, we define a seeds selection optimization problem. To the best of our knowledge, we are the first to study the coverage optimization problem in MCS when the workers are recruited in the social networks.
	\item We propose two algorithms (i.e., \textit{Basic-Selector} and \textit{Fast-Selector}) to select a near-optimal set of users as seeds. \textit{Basic-Selector} adopts an iterative greedy process based on the predicted temporal-spatial coverage, which obtains better coverage but suffers from efficiency drawback when executed on a large-scale social network. \textit{Fast-Selector} significantly reduces the computation by introducing a heuristic utility function, with the observation for interdependency of geographical positions among friends in the online social network. Besides, by further investigating the stop timing of the influence propagation, \textit{Fast-Selector} adaptively adopts different utility functions in various iterations to better maximize the coverage.
	\item We evaluate our approach extensively using two real-world datasets with social network structures (users and their friendships) and users' mobility traces. The experimental results verify that the \textit{Fast-Selector} outperforms baseline methods in obtained temporal-spatial coverage under various settings. Meanwhile, the \textit{Fast-Selector} runs much more efficiently than the \textit{Basic-Selector} while only sacrificing a small proportion of temporal-spatial coverage.
\end{itemize}

\section{Related Work}
In this paper, we leverage the influence propagation in social networks to assist the worker recruitment in MCS. Thus, we review the related literature from two topics: MCS workers recruitment and influence maximization in the social network. 

\subsection{MCS Worker Recruitment}
Recruiting a large group of MCS workers and assigning tasks is a major research topic for MCS.

A set of research works aims at maximizing the data quality of MCS with certain constraints. The authors in \cite{4} propose a novel task allocation framework to maximize the spatial coverage of multiple tasks with the sensing capability constraints. The authors of \cite{5} \cite{7} study develop greedy-based worker selection mechanisms to maximize the temporal-spatial coverage of crowdsensing with the predefined number of workers. Singla et al. \cite{6} propose a novel adaptive worker selection mechanism for maximizing temporal-spatial coverage under total incentive constraint with respect to privacy. \cite{17,18} consider a new version of task assignment problem, where the optimization goal is to maximize the coverage quality under an overall budget constraint. Another set of research studies, one the other hand, aim at minimizing the cost while ensuring a certain degree of the data quality. Zhang et al. \cite{19} and Xiong et al. \cite{8,20} study offline worker selection in piggyback mobile crowd sensing for probabilistic coverage. Karaliopoulos et al. \cite{10} study the user recruitment for mobile crowd sensing over opportunistic networks, in order to minimize the total cost while ensuring the full coverage of POI (points-of-interest). Hachem et al. \cite{11} propose a worker selection framework for MCS, which reduces the number of selected workers by predicting mobile users' future locations in next time slot (or sensing cycle) based on their current location and recent trajectory. Moreover, Liu et al. \cite{13} proposed two task allocation optimization approaches for two different situations. One is when there are few workers and many tasks, and the other is when there are few tasks and many workers. Kandappu et al. \cite{14} designed, developed and experimented with a real-world mobile crowd-tasking platform on the university campus, and study the workers’ behavior and pricing strategies in MCS worker recruitment.

The above mentioned research works all assume that the workers are recruited from a pre-built specialized MCS platforms with an adequately large user group, and their goal is to select an appropriate subset of users as workers. In these studies, the users on the MCS platform are regarded as independent individuals. On the contrary, instead of relying on specialized MCS platforms with large user pools, this paper utilizes social network as the medium to recruit workers.  The potential candidates are not isolated but are connected users on a social network. Therefore, it considers social structures among candidate users, influence propagation model of MCS tasks, and efficient seeds selection algorithm, which are not addressed in the above pieces of literature.

Furthermore, there are a few studies which also attempt to integrate social networking apps into MCS process \cite{35,36,37,38,40}. In these research works, the authors extract various types of useful information (e.g., profile and friendship) from social network to assist the MCS process (e.g., modeling reputation and inferring expertise). These mechanisms and our proposed approach are complementary to each other. For example, these methods can help extract useful social metadata, which may help to extend our propagation models and seed selection process. However, in these studies, MCS task propagation among friends is not introduced to worker recruitment. Moreover, they do not address the seed selection optimization problem with proposed algorithms.

\subsection{Influence Maximization in Social Networks}
The spread of information or influence upon the social network is a fundamental issue in social network analysis. The problem of influence maximization \cite{21} has been well studied as a crucial problem, which can be generalized by finding a subset of influential nodes that can influence the largest number of nodes in the network \cite{22}. Kempel et al. establish that the optimization problem of influence maximization is NP-hard \cite{21}. They use the Greedy algorithm (GA) and prove that the optimal solution for influence maximization can be approximated to within a factor. Authors in \cite{27} propose methods to discover influential event organizers from online social networks who are essential to the overall success of social events. In \cite{28} the authors propose the Comparative Independent Cascade (Com-IC) model that covers the full spectrum of entity interactions from competition to complementarity and study self-influence maximization and complementary-influence maximization problems. The main problem of influence maximization related algorithms is the inefficiency, especially when the social network contains a large number of nodes. Consequently, Chen et al. \cite{24} propose an improved version of the Greedy algorithm, called NewGreedy. To make the result better, it takes the first round with NewGreedy algorithm and the rest rounds using CELF Greedy algorithm, called MixGreedy, which is shown to be more efficient than previously proposed Greedy algorithms. In \cite{25,26} the authors propose a new algorithm called community-based greedy algorithm for mining top-K influential nodes. 

Though the approach proposed in this paper also involves the influence spread on social networks, it differs from the above mentioned studies in the following aspects: 1)\textit{Different optimization goal and constraint.} The optimization goal of above literature is to find a pre-defined number of nodes, with the objective of maximizing the total number of influenced users, while our work is to maximize the temporal-spatial coverage of MCS with one additional constraint (i.e. the maximum number of influenced users). 2)\textit{Different seeds selection algorithm.} Seed selection algorithms in the above literature cannot be directly adopted to solve our problem, because the maximized number of influenced users may not be able to achieve a good temporal-spatial coverage for the MCS task. This is because the influenced users on the social network are always similar in terms of their trajectories \cite{15}, so that sensor readings may be redundant in some subareas while insufficient in others. 3)\textit{Different influence propagation model.} When determining whether the users would be influenced, the adopted models in the above work only consider the influence from their neighbors. In contrast, this paper extends the state-of-the-art propagation models in the social network by introducing MCS-specific features.

\section{system model and problem formulation}

In this section, we first define the system models and present relevant assumptions, including the MCS task specification and influence propagation model on social networks. Then, we formulate the temporal-spatial coverage maximization problem for MCS when the workers are recruited on the social network.

\subsection{MCS Task Specification}
As illustrated in the use case, this paper focuses on the distributed environment monitoring task. We use the temporal-spatial coverage as the metric to characterize the sensing quality, which is widely used in the state-of-the-art work (e.g., \cite{5,6,7}). Specifically, the organizer specifies the target sensing area consisting of a set of subareas, denoted as $S=\{s_1,s_2...s_i...s_m\}$, and sensing duration which is divided into equal-length cycles, denoted as $C=\{c_1,c_2...c_j...c_n\}$. Thus, there are $m\times n$ number of temporal-spatial cells, which forms a set denoted as $S\times C=\{(s_i,c_j)| s_i\in S,c_j\in C\}$. Then, the temporal-spatial coverage is measured as the fraction of temporal-spatial cells being covered by the mobility of a set of workers $W=\{w_1,w_2...w_l\}$, which is denoted as $\frac{|Covered(W)|}{|S\times C|}$, where $Covered(W)$ is the set of temporal-spatial cells with at least one sensor reading.

\subsection{MCS-Specific Influence Propagation Models}
In order to characterize the influence propagation through the social network, there are two mainstream models, i.e., the Independent Cascade model (IC model) and Linear Threshold model (LT model) \cite{21}.  In both models, at any time step, a user is either active (an adopter of the information) or inactive.  In the IC model, when an inactive user becomes active at time $t$, it gets exactly one chance to independently activate its currently inactive neighbors at time $t+1$. In the LT model, the sum of incoming edge weights on any node is assumed to be at most 1 and every user chooses an activation threshold uniformly at random from \cite{1}. At any time-stamp, if the sum of incoming influence (edge weights) from the active neighbors of an inactive user exceeds its threshold, it becomes active. In both models, influence propagates until no more users can become active.

However, it is inappropriate to directly adopt the above models in our MCS worker recruitment problem. This is because, when determining whether the user would be influenced (i.e., accepting the task), both the IC model and LT model only consider the influence from the neighbors in the social network, without taking specific factors about MCS tasks into account.  Thus, in order to model the probability of task acceptance, we extend IC and LT model by integrating MCS-specific factors in Definition 1 and 2, respectively.

\textbf{Definition 1: Extended IC model.} Based on the basic concept of IC model, the probability of user $u$ accepting an MCS task $T$ is calculated as follows:
\begin{equation}
P(T,u)=min\{p_0*I_1(T,u)*I_2(T,u)*\dots*I_L(T,u),1\}
\end{equation}
where $p_0$ is the base influence probability in traditional IC model which is a pre-defined constant, while $I_1(T,u),I_2(T,u),\dots,I_L(T,u)$ are the increase of influence posed by $L$ MCS-specific factors. The intuition behind (1) is that the probability of accepting an MCS task for a certain worker is calculated by aggregating the influence from his/her neighbors and multiple MCS-specific factors. 

\textbf{Definition 2: Extended LT model.} Based on the basic concept of LT model, the threshold of each inactive user is defined as follows: 
\begin{equation}
\theta(T,u)=\frac{\theta_0}{I_1(T,u)*I_2(T,u)*\dots*I_L(T,u)}
\end{equation}

where $\theta_0$ is the base influence threshold which is a pre-defined constant, while $I_1(T,u),I_2(T,u),\dots,I_L(T,u)$ are the increase of influence (or the decrease of influence threshold) posed by $L$ MCS-specific factors. The intuition behind (2) is that the threshold of accepting an MCS task is influenced by the original threshold of LT model and multiple MCS-specific factors.

In both the extended IC model and LT model, the key issue is how to choose the influence increase function $I_1(T,u),I_2(T,u),\dots,I_L(T,u)$. According to Weber–Fechner law in the field of psychophysics \cite{39}, with the increase of external stimulus, humans’ perception is enhanced but the degree of enhancement decreases (i.e., diminishing return property). To keep this property, we choose function $I(x,I_{max})=(I_{max}-1)\sqrt{1-(1-x)^2}+1$ to measure the increase of influencing probability, where $x$ is the input probability increasing parameter varying from one factor to another and $I_{max}$ is the maximum increase. Then we have $I(0,I_{max})=1$, $I(1,I_{max})=I_{max}$, $\frac{\partial I(x,I_{max})}{\partial x}>0$ for $x\in(0,1)$.This function is borrowed from \cite{31}, in which it is used to measure the probability increase for attending certain social-network-based events.

There are multiple MCS-specific factors, and we illustrate how the input probability increasing parameter $x$ in $I(x,I_{max})$ is defined in two exemplary factors: (a) whether the incentive is attractive; (b) whether the topic of a task is interesting. We choose these two because they are significant factors to influence the users’ participation willingness according to the state-of-the-art studies \cite{30, 31}.

\textbf{Factor Example 1: Topical Interest.} We use $I_1(T,u)$ to measure the influence increase brought by the match between the user's interest and the topic of the MCS task.

\begin{equation}
I_1(T,u)=I(Cos(\overrightarrow{T.topic},\overrightarrow{u.interest}),I_{max1})
\end{equation}

where $I_{max1}$ is the upper bound of increase which is a pre-defined constant, and $Cos(\overrightarrow{T.topic},\overrightarrow{u.interest})$ is the cosine similarity between vector $\overrightarrow{T.topic}$ and $\overrightarrow{u.interest}$.

For example, assuming that there is a full set containing five topics, i.e., \{"air quality", "environment monitoring", "sports", "movie", "politics"\}, we give each topic an index. If the $\overrightarrow{T.topic}$ in the above use case application AirSense is \{ "air quality", "environment monitoring"\} and $\overrightarrow{u.interest}$ of a specific user $u$ is \{"air quality", "sports", "movie"\}, then $\overrightarrow{T.topic}$ =(1,1,0,0,0) and $\overrightarrow{u.interest}$ = (1,0,1,1,0).

\textbf{Factor Example 2: Incentive Attraction.} The influence increase brought by the match between the task's provided reward and the user's expectation is denoted as $I_2(T,u)=I(F(T,u),I_{max2})$, where $I_{max2}$ is the upper bound of increase pre-defined as a constant and $F(T,u)$ is the attraction of task $T$'s incentive for $u$. For example, $F(T,u)$ can be defined as: 
\begin{equation}
\begin{split}
&F(T,u) = tanh(T.incentive-u.minimum)\\
&  if\ u.minimum\leq T.incentive\\
&  F(T,u) = 0\ if\ u.minimum> T.incentive
\end{split}
\end{equation}

where $u.minimum$ is lower bound incentive reward influencing $F(T,u)$. The intuition behind (4) is that: if the reward is lower than the user’s lowest expectation, then the reward cannot increase the probability of acceptance (i.e., $F(T,u)=0$, thus $I_2(F(T,u),I_{max2})=1$). Otherwise, it increases with the increase of the provided incentive reward, but the maximum value of $F(T,u)$ is 1 (i.e., $I_2(T,u)=I_{max2}$).

\subsection{Problem Definition}

Based on the above MCS task specifications and extended influence propagation models, we define the worker recruitment problem on MSN as follows:

Given a set of temporal-spatial cells $S\times C=\{(s_i,c_j)| s_i\in S,c_j\in C\}$, a set of mobile users $U=\{u_1,u_2...u_k\}$ on the social network who have at least a certain number of check-in data (including the timestamp and location) within $S\times C$, we denote the set of nodes $U^{'}\subseteq U$ selected from the social network as initial seeds. $f(U^{'})\subseteq U$ is the set of influenced users (i.e., recruited workers) with the above extended IC or LT model on $U^{'}$. We further denote $Covered(f(U'))$ as the set of temporal-spatial cells covered by the mobility of $f(U^{'})$. Then, the problem is then to find $U^{'}\subseteq U$, with the objective to
$$
Maximize\ \frac{|Covered(f(U^{'}))|}{|S\times C|}
$$
$$ 
Subject\ to\ |U^{'}|\leq p\ and\ |f(U^{'})|\leq q
$$
where $p$ is the maximum number of seeds and $q$ is the maximum number of recruited workers.
\section{Basic-selector}

\subsection{Data Preparation and Mobility Profiling}

In terms of how the MCS task is performed, we utilize the piggyback MCS manner \cite{32}, which intends to reduce energy consumption by leveraging smartphone application opportunities to execute sensing tasks and report sensed result. Thus, in this paper, we assume that the mobile device would collect and report the sensing data when workers use their MSN applications to check-in. Given the check-in data of mobile users on the social network, we compute the check-in/mobility profile of each user, i.e., the probability of each user performing at least one check-in at a particular subarea in a given sensing cycle. 

Specifically, it computes the profile of each user with following two steps: We first map each worker's historical check-in traces onto sensing cycles. Then we count the average number of check-in behaviors by each worker  at each subarea $s_i$ in each cycle $c_j$, which is denoted as $\lambda_{w,i,j}$. For example, to estimate  $\lambda_{w,i,j}$ for sensing cycle $c_j$ from 08:00 to 09:00 of a specific day, we count the average number of check-in by  at $s_i$ during the same period in the historical records. 

Similar to several state-of-the-art studies \cite{17,19,29}, we assume that the check-in behavior follows an inhomogeneous Poisson process. Thus, the probability of a worker  to perform check-in for $h$ times at subarea $s_i$ in sensing cycle $c_j$ can be modeled as:
\begin{equation}
Pro_{i,j}(w,h)=\lambda_{w,i,j}^{h}*e^{-\lambda_{w,i,j}}/{h!}
\end{equation}
Therefore, we can estimate the probability of worker  perform at least one check-in during cycle $c_j$ at $s_i$ as follows:
\begin{equation}
Pro_{i,j}(w)=\sum_{h=1}^{\infty}Pro_{i,j}(w,h)=1- e^{-\lambda_{w,i,j}}
\end{equation}
Since we assume that the workers of MCS task would collect and report the sensing data when using their MSN applications for check-in, the probability of worker  providing at least one sample for the assigned task during cycle $c_j$ at $s_i$ is predicted as:
\begin{equation}
\alpha_{i,j}(w)=1- e^{-\lambda_{w,i,j}}
\end{equation}

\subsection{Basic Seeds Selection Algorithm}
Given users’ check-in/mobility profile and the influence propagation models, we propose a greedy-based seed selection algorithm named \textit{Basic-Selector}, which iteratively selects the “most beneficial” seeds. Here, the “benefit” (or the utility) is defined as the temporal-spatial coverage increase when adding $u_x$ into the seeds set $U'$ (see formula (8)). The algorithm keeps selecting and adding new users until the number of seeds reaches to the limitation or none of the unselected users can increase the estimated temporal-spatial coverage

\begin{equation}
Utility(u_x)=|Covered(f(U'\cup \{u_x\}))|-|Covered(f(U'))|.
\end{equation}

\begin{figure}[!t]
    \centering \includegraphics[width=3in]{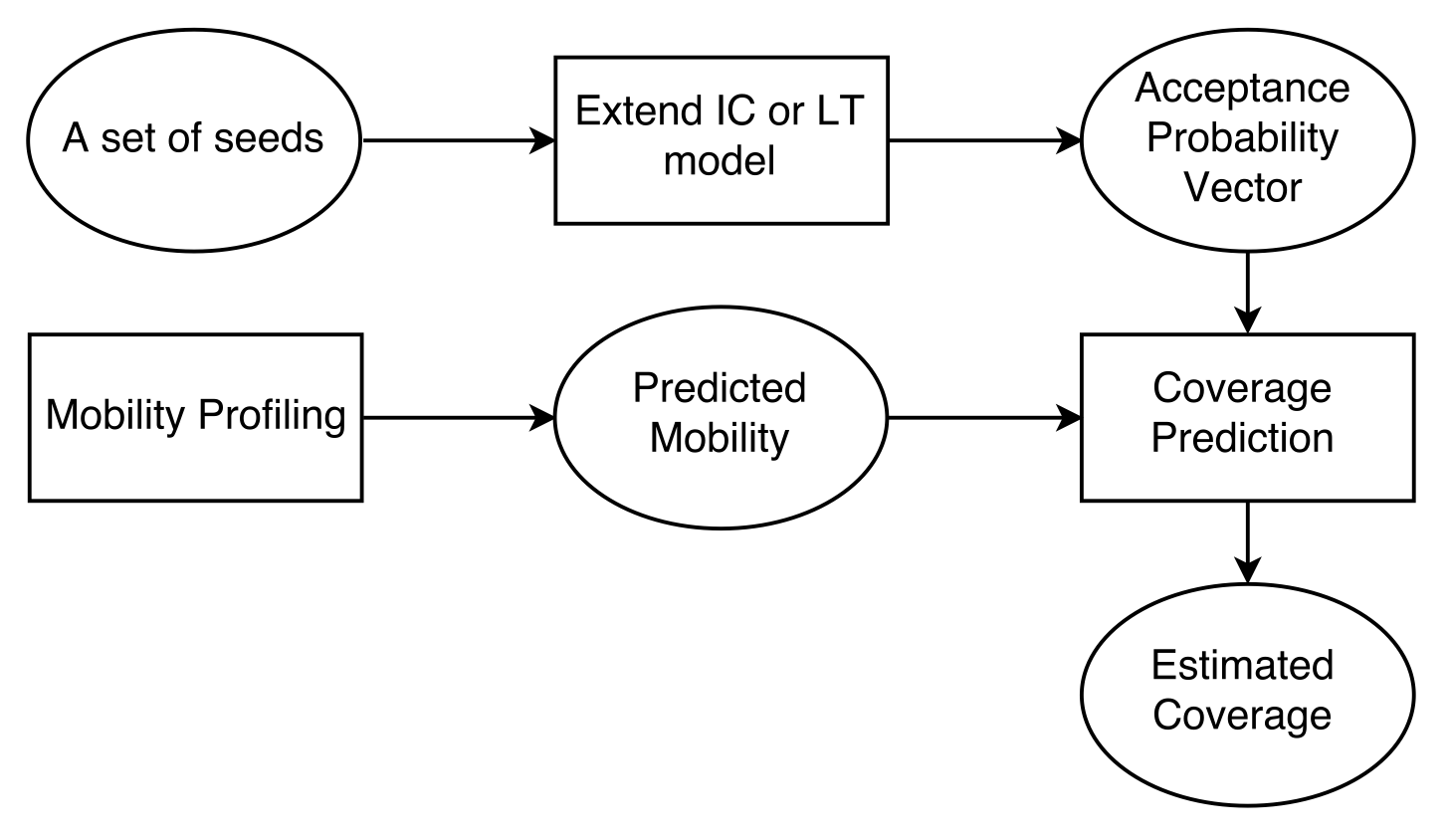}
    
    \caption{ Temporal-spatial coverage estimation process adopted in \textit{Basic-Selector} algorithm \label{fig_1}}
\end{figure}

The temporal-spatial coverage of finally recruited workers achieved by a given seeds set $U^{'}$ is estimated as following steps (see Fig. ~\ref{fig_1}).

First, we calculate the probability of each user to become the finally recruited worker. Inspired by the existing influence maximization approaches \cite{22,26}, we run Monte-Carlo simulations \cite{23} of the extended influence propagation model for sufficiently many times (typically 1000) to obtain an accurate estimate of the influence spread. The probability of a given user $u_x$ to become the worker is denoted as $P(T,u_x)=p_0*I_1(T,u_x)*I_2(T,u_x)$. Thus, at the end of this step, the probability of every user on the social network form a vector, which is called acceptance probability vector. The vector is formally denoted as $<P(T,u_1), P(T,u_2)......P(T,u_k)>$.

Then, given a specific temporal-spatial cell $(s_i,c_j)$ for task $T$, we calculate its probability to be covered as follows:
\begin{equation}
\varPhi(i,j)=1-\prod_{u_z\in U^{'}}(1-\alpha_{i,j}(u_z) * P(T,u_z))
\end{equation}
Finally, the overall temporal-spatial coverage is estimated as: 
\begin{equation}
Covered(U^{'})=\frac{\sum_{i=1}^{n}\sum_{j=1}^{m}\varPhi(i,j)}{m*n}
\end{equation}

Although the \textit{Basic-Selector} is easy to implement, it suffers from the following severe drawbacks in efficiency: (i) The Monte-Carlo simulations that are executed sufficiently many times (typically 1000) to obtain an accurate estimate of spread prove to be very expensive \cite{24}. (ii) The algorithm makes O(kp) calls to the spread estimation procedure (Monte-Carlo simulations in this case) where k is the number of nodes in the graph and p is the size of the seed set to be picked. To further investigate the inefficiency concern, we have designed the following empirical study.

\begin{figure}[!t]
	\centering \includegraphics[width=3in]{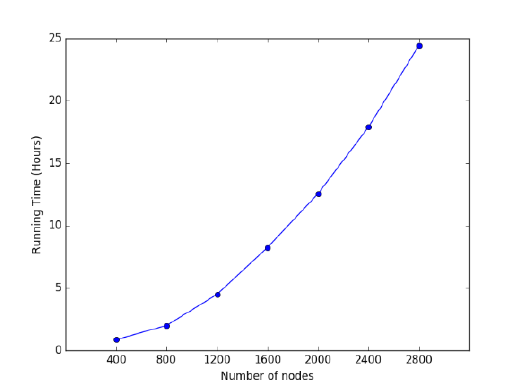}
	
	\caption{Running time for \textit{Basic-Selector} \label{fig_2}}
\end{figure}

First, we conduct the data profiling and mobility profiling based on the Brightkite dataset, which is a real-world MSN dataset. Then, the \textit{Basic-Selector}  is executed on a standard commercial sever by varying the number of users on the social network\footnote{\textit{To vary the number of users, we randomly select a different number of users on the dataset, on which the Basic-Selector is executed.}}. Fig. ~\ref{fig_2} shows the computation time with a different number of users. From the figure, we can see that the computation time increases dramatically with the increase in the number of users. Moreover, we can see that even finding a small seed set (25 in this empirical study) in a moderately large network (e.g. 2800 users) could take a day to complete on a modern server machine, whose configuration is presented later in the experimental evaluation section. Therefore, the inefficiency concern makes \textit{Basic-Selector} inappropriate to be adopted on a large-scale social network.

\section{Fast-selector}
\subsection{Motivation and Intuitions}

From the analysis and empirical study in Section 4.2, we can see that \textit{Basic-Selector} has drawback in its efficiency. Thus, to make our proposed MCS worker recruitment mechanism applicable in large-scale social networks, we need to design a more efficient seed selection algorithm.

Actually, the computation time of \textit{Basic-Selector} is mainly consumed in the multiple rounds of Monte-Carlo simulations while invoking the influence propagation model for each given set of seeds. To avoid running such time-consuming simulations, we attempt to estimate the benefit of seeds through an alternative way. Specifically, we introduce an intuition-based metric to measure the utility when adding one user into the seeds set. The intuition mainly consists of the following two aspects. 

First, in order to maximize the temporal-spatial coverage of an MCS task, we intuitively prefer to select more "influential" users, that is, nodes being able to propagate the MCS task to a larger number of users on the network. As the degree of a user on the social network graph is a widely-used indicator for measuring user's influence \cite{24}, \textit{our algorithm prefers to select users with a higher degree as the seeds.} Here the degree refers to the number of followers in a directed graph (e.g., Twitter), while it referring to the number of friends in an undirected graph (e.g., Facebook). 

Second, we also note that our temporal-spatial coverage maximization problem is different from the influence maximization. As described in the related work, the goal of influence maximization problem is to maximize the number of influenced users \cite{21}, while ours is to optimize temporal-spatial coverage of an MCS task achieved by influenced users. The maximized number of influenced users may not be able to achieve an optimal temporal-spatial coverage. This is because a user commonly spreads his/her influence to others in the same community so that the influenced users obtained by influence maximization algorithms commonly share similar routine trajectories \cite{15,26}. Thus, even if the number of workers is maximum, the sensor readings of an MCS task may be redundant in some subareas while insufficient in others, which has a negative impact on the coverage maximization. Motivated by the above observation, we have another intuition, that is, if the selected seeds are distributed more uniformly within the entire sensing area, then the influenced users (i.e., recruited workers) would be distributed more uniformly as well, which would lead to higher temporal-spatial coverage. As a result, \textit{we also intuitively prefer to select users that can make the mobility trace of selected seeds more diverse after adding it.}

\subsection{Na\"{\i}veFast Algorithm}
According to the above two intuitions, we propose a rank utility function as $R(u)=\beta * DegreeRank(u,A)+(1-\beta )*TriDiffRank(u,A)$ for adding one user $u\in U/A$ into the seed set $A$, where $A$ is the seeds set obtained in the previous iteration. $DegreeRank(u,A)$  is the rank of user $u$ in terms of degree among the candidate nodes (i.e., $U/A$), where the users with the bigger degree have bigger rank number $DegreeRank(u,A)$. $TriDiffRank(u,A)$ is the rank of user $u$ in terms of average trajectory difference with the nodes in $A$, where the users with smaller average trajectory difference have bigger rank number $TriDiffRank(u,A)$. $\beta$ is a parameter used to balance $DegreeRank(u,A)$ and $TriDiffRank(u,A)$. Specifically, the average trajectory difference between $u$ and the nodes in $A$ is calculated by using \textit{Cosine Similarity}. As we declared in section 4.1,  $\lambda_{w,i,j}$ is the average number of check-in data by each worker $w\in W$ at each subarea $s_i$ in each cycle $c_j$. For each work $w$, we define the $m\times n$ mobility matrix as $M_w$, where the element in row $i$ and column $j$ is assigned as $\lambda_{w,i,j}$. Then we connect each row of matrix $M_w$  into a vector and represent it as:
$$
M_w=[(\lambda_{w,1,1},...,\lambda_{w,1,n}),...,(\lambda_{w,m,1},...,\lambda_{w,m,n})]
$$
Given a specific user $u$ and seed set $A$ obtained in the previous iteration, then the average trajectory difference between user $u$ and users in $A$ can be calculated by the following equation:
\begin{equation}
\frac{\sum_{x\in A}\frac{M_x\cdot M_u}{||M_x||||M_u||}}{|A|}
\end{equation}
where $\frac{M_x\cdot M_u}{||M_x||||M_u||}$ is the cosine similarity between $M_x$ and $M_u$ and $|A|$ is the number of users in $A$. The smaller the (11) is, the bigger the average trajectory difference between $u$ and users in $A$ is.

A simple and straightforward algorithm is that in each iteration, it selects the user with maximum rank-based utility and add it into the seeds set, until the size of seeds reaches to the limitation. It is referred to as \textit{Na\"{\i}veFast} in this paper, which serves as a baseline method in the latter experimental study.  \textit{Na\"{\i}veFast} is much more efficient than the above \textit{Basic-Selector} because it avoids considerable computation on Monte-Carlo simulations of the influence propagation model in all iterations.

\subsection{Fast-Selector: Two-phase Approach}

The \textit{Na\"{\i}veFast} algorithm also has its drawback, when we investigate the complex propagation process of influence model. During the iterative greedy selection process, iterations can be divided into two different phases in terms of the timing when the MCS task propagation stops. In the first phase, when the number of selected users is small, the propagation will stop when there is no user to be influenced any longer. During this phase, the number of finally recruited workers would not reach to the limitation so that we call this phase "budget-insensitive". In the second phase, with the increase in the number of selected users, the MCS task propagation stops when the number of finally recruited workers reaches to the limitation so that we call this phase "budget-sensitive". Different phases have the following different properties.

\begin{itemize}
	\item In the \textit{budget-insensitive phase}, when adding a new user into the seeds set, the overall obtained temporal-spatial coverage will increase. This is because, in this phase, the influenced users by the seeds set obtained in the $i$th iteration will also be influenced by the seeds set obtained in the $i+1$th iteration.
	\item In the \textit{budget-sensitive phase}, however, the influenced users by the seeds set obtained in the $i$th iteration may not be influenced by the seeds set obtained in the $i+1$th iteration. Thus, in this phase, if we still iteratively add a new user into the seeds set, the obtained temporal-spatial coverage may not necessarily increase.
\end{itemize}

Based on the above characteristics, we propose the following two-phase-based fast seeds selection algorithm named \textit{Fast-Selector}. By combining the idea of \textit{NaiveFast} and \textit{Basic-Selector}, different approaches and utility functions are adaptively used in different phases. In the budget-insensitive phase, we select users iteratively based on the intuition-based rank utility function $R(u)$, which is the same as \textit{Na\"{\i}veFast}. In the budget-sensitive phase, as the propagation process becomes more complicated, we need to adopt more sophisticated utility function for making a more cautious decision about which should be the next newly added seed. Specifically, we choose the approach and utility function in (8) and (10) of \textit{Basic-Selector} to iteratively select the user, until the number of seeds reaches to the limitation or none of the unselected users can increase the estimated temporal-spatial coverage. The pseudocode of Fast-Selector is shown in Algorithm ~\ref{alg:FS}, where the variable \textit{flag} indicates if the seeds selection is in the budget-sensitive phase or budget-insensitive phase. The value of the $flag$ is obtained through the function $networkSpread(U^{'}, q)$, which simulates the propagation using either extend IC or LT model described in section 3. 

\renewcommand{\algorithmicrequire}{\textbf{Input:}}
\renewcommand{\algorithmicensure}{\textbf{Output:}} 
\begin{algorithm}[htb]         %算法的开始
	\caption{Fast-Selector}             %算法的标题
	\label{alg:FS}                  %给算法一个标签，这样方便在文中对算法的引用
	\begin{algorithmic}[1]                %不知[1]是干嘛的？
		\REQUIRE ~~\\                          %算法的输入参数：Initialization
		the users set $U$, the maximum number of seeds $p$, the maximum number of workers $q$.
		\ENSURE ~~\\      
		the selected seed set $U^{'}$.
		%Ensemble of classifiers on the current batch,  $E_n$;
		
		\STATE set $U^{'}=\emptyset$ and $flag=false$.
		
		\WHILE {$|U^{'}|<p$ and $flag=false$}
			\STATE select $u$ from $U$ with maximum  $R(u)$
			\STATE $U^{'} = U^{'} \cup \{u\}$
			\STATE $U=U-\{u\}$
			\STATE $flag=networkSpread(U^{'}, q)$
		\ENDWHILE
		
		\IF {$flag=True$}
		
		\WHILE {$|U^{'}|<p$}
			\STATE use \textit{Basic-Selector} to select user $u$ from $U$
			\IF {coverage does not increase for any $u$}
				\STATE break
			\ELSE
				\STATE $U^{'} = U^{'} \cup \{u\}$
				\STATE $U=U-\{u\}$
			\ENDIF
		\ENDWHILE
		
		\ENDIF
		\RETURN $U^{'}$
	\end{algorithmic}
\end{algorithm}

\section{Evaluation}
In this section, we report the evaluation results using two large-scale real-world MSN datasets to verify the effectiveness of our approach, which mainly consists of the following components. First, we present the datasets, basic experiment settings, and baseline methods. Second, the detailed evaluation results with respect to the \textit{Fast-Selector} and baseline methods are presented and compared when the size of nodes is relatively big. Third, we evaluate \textit{Fast-Selector} and \textit{Basic-Selector} through the comparative study when the size of nodes is relatively small. Finally, we summarize the above experimental results to get conclusions or implications.

\subsection{Datasets Description and Baseline Methods}

When selecting the dataset for evaluation, there are two criterions. First, it should contain the mobility traces from a large number of users, which will be used for mobility perdition and coverage estimation. Second, it should contain the friendship network among those users, which will be used in the seed selection algorithm and the simulation of MCS task propagation on social networks. Although there are several other open datasets containing mobility traces (such as the dataset used in \cite{17,38}), there are relatively few containing both the mobility trace and friendship network structure. Thus, we use two open datasets (i.e., Brightkite and Gowalla) in our experiments, because they contain both users' locations and friendship network.

\begin{figure}[!t]
	
	\centering \includegraphics[width=3in]{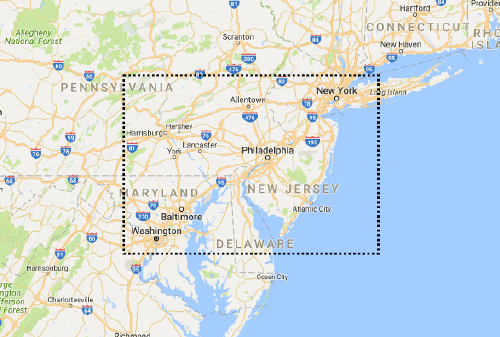}
	
	\caption{Entire sensing area (subareas in the sea are removed) \label{fig_4}}
\end{figure}

When a user checks in, an individual record formatted as $<$user-id, check-in-time, latitude, longitude, location-id$>$ is produced. It is difficult to detect users' movements when the distances of users' movements are small and the scale of the physical world is large. Hence, to make sure the users' movements are correctly detected, we employ part of the original Brightkite and Gowalla datasets. In the employed two subsets, the users are distributed in 240km$\times$200km rectangle regions except those in the sea (see Fig. ~\ref{fig_4}). Besides, we assume that the entire sensing area is divided into 480 equal-length virtual subareas (i.e., 10$km$$\times$10$km$ per subarea). By removing the subareas in the sea, there are 344 subareas which we considered in the experiment including New York, Washington, and Philadelphia where users are densely distributed. For the Brightkite dataset and Gowalla dataset, the investigated check-in records are over the period of April 2008 through October 2010 and December 2009 through October 2010, respectively. The details of the datasets are summarized in Table~\ref{table_1}.

\begin{table}[!t]
	\renewcommand{\arraystretch}{1.3}
	\caption{Datasets Summary}
	\label{table_1}
	\centering
	\begin{tabular}{c||c||c}
		\hline
		\bfseries \  & \bfseries Brightkite & \bfseries Gowalla \\ 
		\hline\hline
		\#users & 8650 & 10693 \\
		\hline
		\#edges & 32536 & 55506 \\
		\hline
		\#check-in & 458648 & 333915 \\
		\hline
		average check-in & 53.023 & 31.227 \\
		\hline
	\end{tabular}
\end{table}

Then, all these data records are reselected randomly every week as a testing dataset, while the rest are used as training dataset. Specifically, we define that sensing duration is 8:00 am$\sim$6:00 pm on each day and each sensing cycle last 2 hours, so that the total sensing duration consists of $5\times7=35$ cycles in each testing dataset. After the datasets are prepared, our experiment uses the records in the training dataset for seeds selection, and we tested the coverage using the testing data records. Finally, the results are regarded as the average performance.

We provide the following baseline seeds selection methods for comparative studies.
\begin{itemize}
	\item \textit{MaxDegree} - This method adopts a greedy algorithm, and it incrementally selects users with the maximum degree on the social network, until the number of seeds reaches to the limitation. 
	\item \textit{MaxCov} - This method selects a limited number of users with the maximum temporal-spatial coverage as the seeds. Here, we adopt the algorithm proposed in \cite{19} to implement this process. 
	\item \textit{Heuristic Greedy (HG)} - This method adopts a greedy algorithm, in which it incrementally selects users with highest heuristic utility. The heuristic utility is defined as $Mobility(u)*Degree(u)$, where $Mobility(u)$ is the number of subareas $u$ has passed according to the historical records, while $Degree(u)$ is the degree of $u$ in the social network.
	\item \textit{Na\"{\i}veFast} - This method adopts similar greedy process as \textit{Fast-Selector}. The difference is that it only uses the intuition-based rank utility $\beta * DegreeRank(u,A) + (1-\beta)*TrDiffRank(u,A)$ to select seeds until the number of seeds reaches to the limitation. Different from \textit{Na\"{\i}veFast}, \textit{Fast-Selector}	adopts different utility functions in budget-insensitive and budget-sensitive phase.
\end{itemize}

We use two metrics to evaluate performance, namely the temporal-spatial coverage and running time. We carried out experiments using a server with 32 Intel Xeon(R) E5 CPU and 128GB memory. \textit{Basic-Selector}, \textit{Fast-Selector}, and other baseline algorithms were implemented in Java. 

\subsection{Experimental Setups}
The parameters used in the experiment are set as follows. 

First, we evaluate the effect of parameters on performance, including the limited number of seeds and recruited workers. We vary the number of seeds from 25 to 100 while fixing the number of recruited workers as 2000 or 5000, respectively. Second, for a given limited number of seeds and recruited worker, there are also some other parameters, in terms of the extended IC model and LT model, need to be set up. As varying each parameter is too time-consuming, we randomly set their values and run 20 rounds of experiments or set a fixed one. The average performance is calculated as the experimental results demonstrated latterly. To determine the range or the fixed value, we take the parameter setting of previous work on influence maximization \cite{21,24,26,31} and our actual network into consideration, and randomly set the influence probability $p_0$ by neighbors of IC model as [0.1, 0.5] and the basic threshold $\theta_0$ of LT model as [0.5, 0.9]. Third, we set the $F(T,u)$ and $Cos(\overrightarrow{T.topic},\overrightarrow{u.interest})$ based on a survey \cite{34}. In this survey, we give five typical MCS tasks with description. Each task aims at collecting one type of sensing data, including the air quality, traffic congestion status, noise level, missing manhole cover, and the flow of people. We recruited 93 volunteers (50 males, 43 females, and aged from 20 to 58) through online advertisements sent to about 1,000 people working or studying across six provinces in China. Each of the recruited volunteers are asked to answer 10 questions in the survey regarding their interest level and expectation of the minimum rewards towards each task. Based on the survey results, we simulate the distribution of two parameters (i.e., $F(T,u)$ and $Cos(\overrightarrow{T.topic},\overrightarrow{u.interest})$), and then generate the value of each worker according to this distribution. Finally, $\beta$ is a parameter used to balance $DegreeRank(u,A)$ and $TrDiffRank(u,A)$. The parameter $\beta$ in \textit{Na\"{\i}veFast} and \textit{Fast-Selector} is set at 0.64 for Brightkite and 0.56 for Gowalla since it usually achieves the highest coverage according to our experiments. Table~\ref{table_2} summarizes the parameters settings in our experiment. 

\begin{table}[!t]
	\renewcommand{\arraystretch}{1.3}
	\caption{Parameter Settings}
	\label{table_2}
	\centering
	\begin{tabular}{c||c}
		\hline
		\bfseries Parameters & \bfseries Settings\\
		\hline\hline
		$p$ & 25, 50,75,100\\
		\hline
		$q$ & 2000, 5000\\
		\hline
		$p_0$ & Randomly generate from [0.1,0.5] \\
		\hline
		$\theta_0$ & Randomly generate from [0.5,0.9] \\
		\hline
		$I_{max}$ & $I_{max1}=3$, $I_{max2}=1.5$ \\
		\hline
		$F(T,u)$ & Generate based on the survey \\
		\hline
		$Cos(\overrightarrow{T.topic},\overrightarrow{u.interest})$ & Generate based on the survey \\
		\hline
		$\beta$ & Brightkite: 0.64, Gowalla:0.56 \\
		\hline
	\end{tabular}
\end{table}

\subsection{Performance Comparison}
This experiment is to compare \textit{Fast-Selector} with other baseline methods using different influence propagation models (extended IC and LT model), in different datasets (Brightkite and Gowalla), and under various settings (the number of seeds and recruited workers). The parameters of interest and incentive are set differently for each MCS task, but we only report the average results due to the space limit.

\begin{figure}
	\subfigure
		\centering \includegraphics[width=0.24\textwidth]{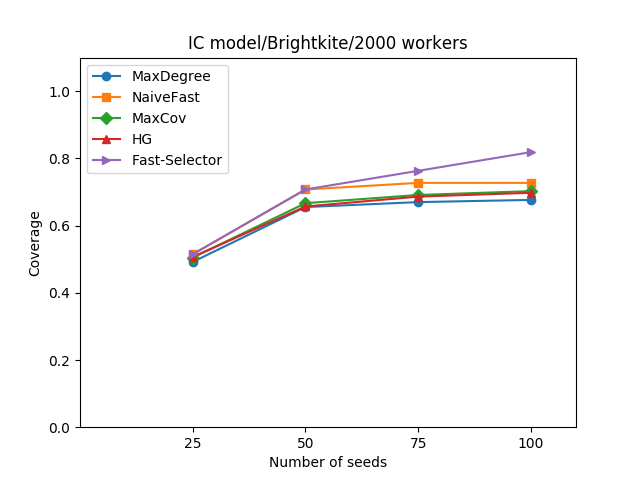}
		\includegraphics[width=0.24\textwidth]{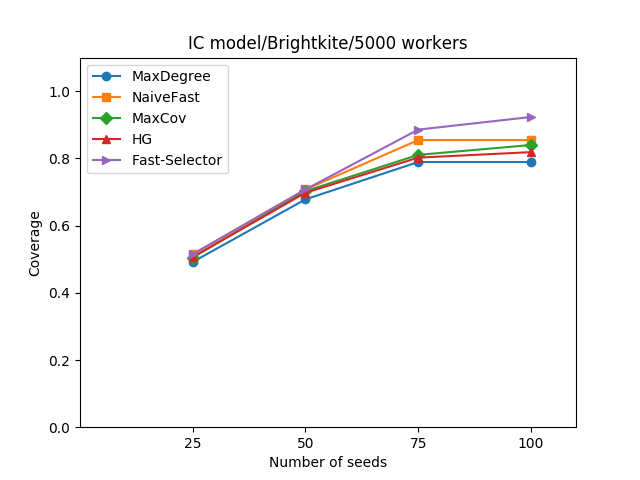}
	\subfigure
	
	\subfigure
		\centering \includegraphics[width=0.24\textwidth]{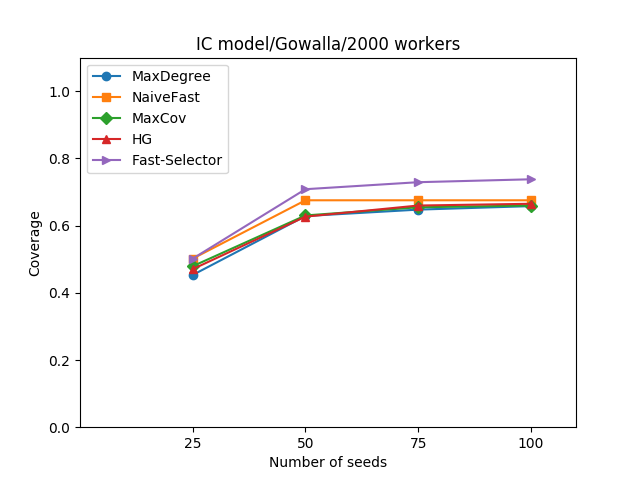}
		\includegraphics[width=0.24\textwidth]{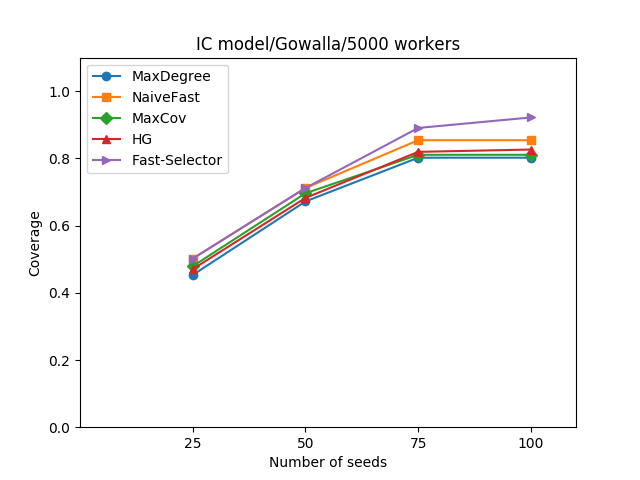}
	\subfigure
	
	\subfigure
	\centering \includegraphics[width=0.24\textwidth]{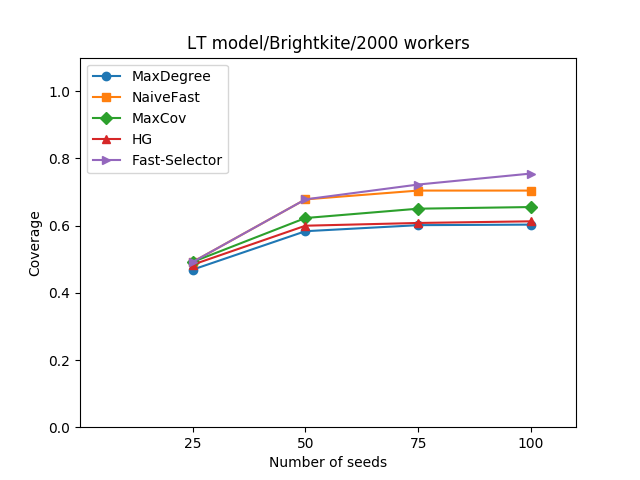}
	\includegraphics[width=0.24\textwidth]{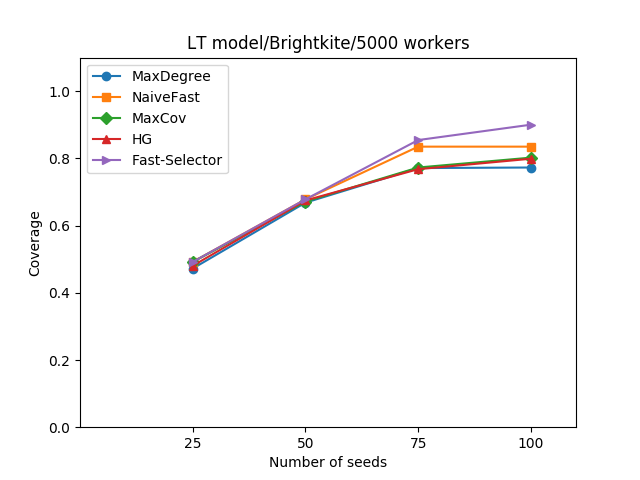}
	\subfigure
	
	\subfigure
	\centering \includegraphics[width=0.24\textwidth]{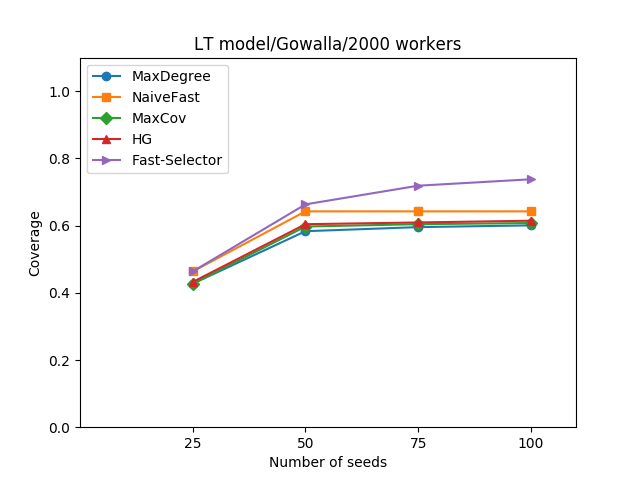}
	\includegraphics[width=0.24\textwidth]{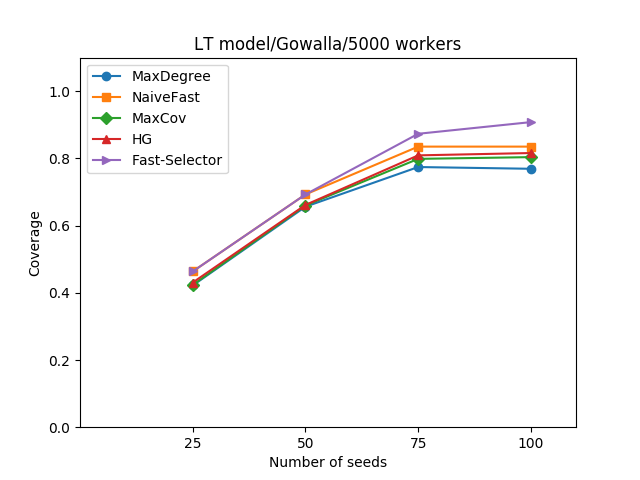}
	\subfigure
	
	\caption{Coverage comparison for different models and datasets (under various settings in number of seeds and workers) \label{fig_5}}
\end{figure}

From Fig. ~\ref{fig_5} we can see that the \textit{Fast-Selector} performs significantly better than the \textit{MaxDegree} and \textit{MaxCov} by achieving higher temporal-spatial coverage in different datasets or using different influence propagation models. For both two datasets and two influence propagation models, the \textit{Fast-Selector} performs better than the \textit{Na\"{\i}veFast} in some settings (e.g., IC model/Brightkite/2000 workers, 75 or 100 seeds), while performs almost equally in other settings (e.g., IC model/Brightkite/2000 workers, 25 or 50 seeds). By re-checking the running log, we found that the equal performance happens when the seeds selection process of \textit{Fast-Selector} stops before the utility function is switched from budget-insensitive phase to budget-sensitive phase. In such cases, it degenerates to the \textit{Na\"{\i}veFast}. Otherwise, \textit{Fast-Selector} outperforms \textit{Na\"{\i}veFast}. In summary, under various settings, \textit{Fast-Selector} averagely outperforms the \textit{MaxDegree}, \textit{HG}, \textit{MaxCov} and \textit{Na\"{\i}veFast} (in terms of coverage) by 12.9\%, 9.6\%, 8.0\% and 4.4\%. For instance, for a sensing target area with 40 subareas (5$km$$\times$5$km$ per subarea), \textit{Fast-Selector} can cover more than 129 $km^{2}$ and 44 $km^{2}$ per time slot than \textit{MaxDegree} and \textit{Na\"{\i}veFast}, respectively.

\begin{figure}[!t]
	\centering \includegraphics[width=3.8in]{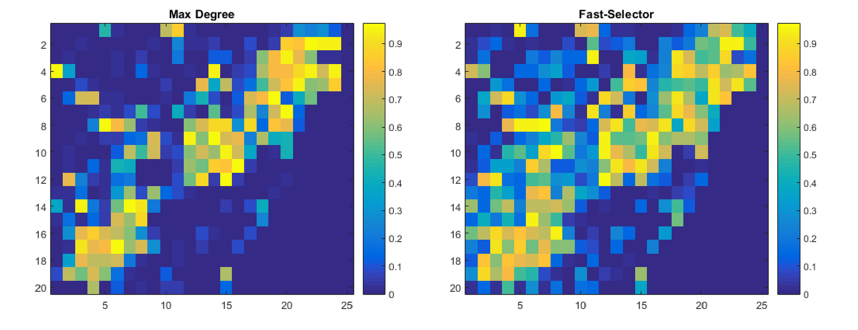}
	
	\caption{Percentage of covered temporal-spatial cell for each subarea (left: \textit{MaxDegree}, right: \textit{Fast Selector}) \label{fig_6}}
\end{figure}

As introducing the trajectory difference is an important intuition for \textit{Fast-Selector}, we further investigate how it helps to optimize the coverage by printing out the coverage status of each temporal-spatial cell between \textit{Fast-Selector} and \textit{MaxDegree}. Fig. ~\ref{fig_6} shows the percentage of the covered cell for each subarea during the 35 sensing cycles. From the results, we can see that the subareas with higher coverage percentage by \textit{MaxDegree} tend to be concentrated, which confirms the aforementioned observation that a user commonly spreads his/her influence to others within the same community with similar routine trajectories. On the contrary, \textit{Fast-Selector} considers the trajectory difference to make the recruited workers distributed more uniformly, which is beneficial for the coverage maximization. The subareas in the bottom-right of the figure fall into the sea (see Fig. ~\ref{fig_4}) without workers passing by, which have already been removed from the experiment.

\begin{figure}
	\subfigure
	\centering \includegraphics[width=0.24\textwidth]{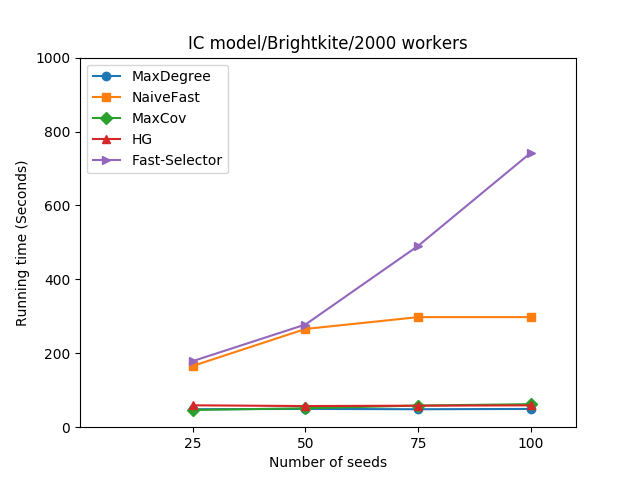}
	\includegraphics[width=0.24\textwidth]{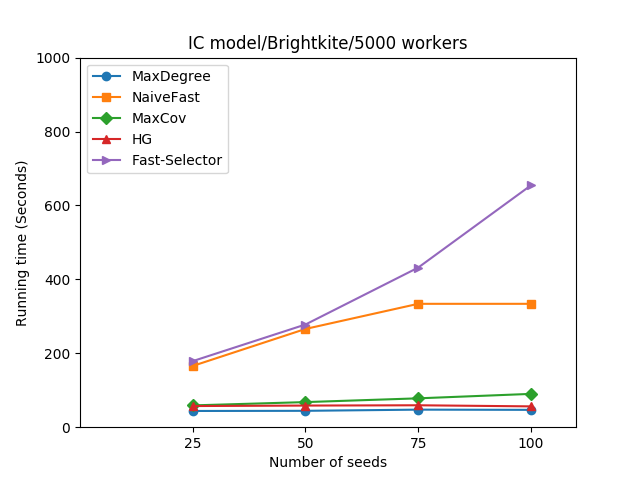}
	\subfigure
	
	\subfigure
	\centering \includegraphics[width=0.24\textwidth]{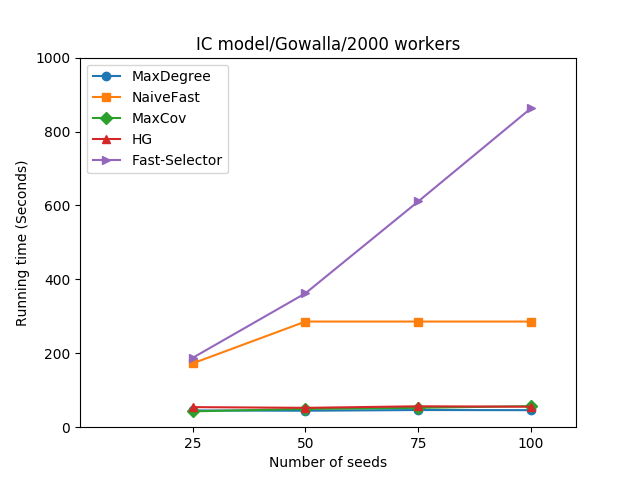}
	\includegraphics[width=0.24\textwidth]{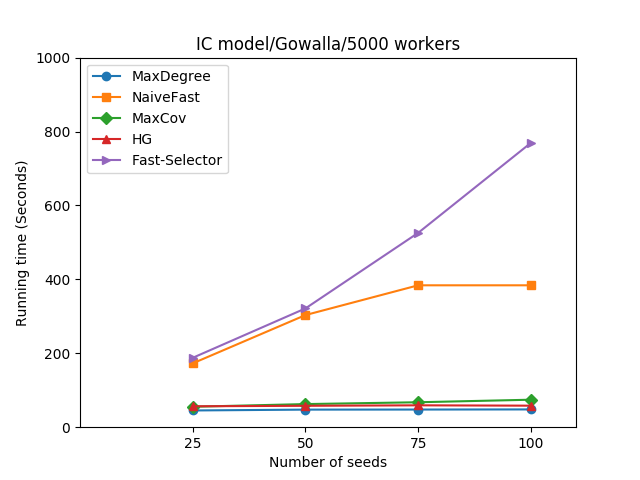}
	\subfigure
	
	\subfigure
	\centering \includegraphics[width=0.24\textwidth]{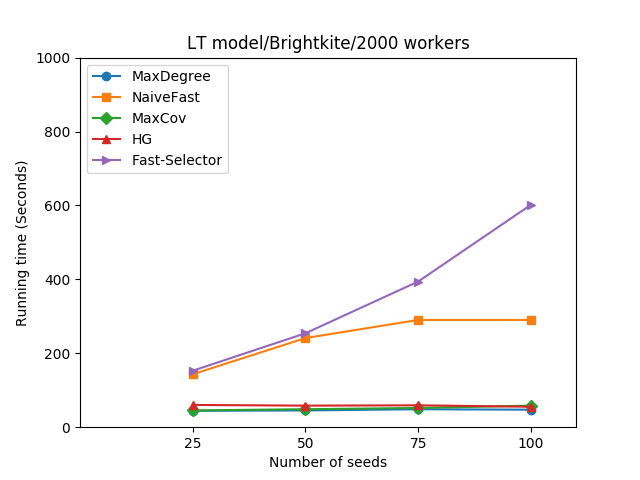}
	\includegraphics[width=0.24\textwidth]{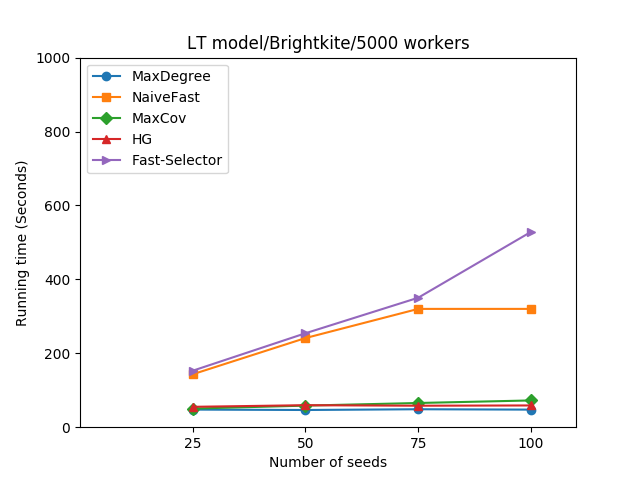}
	\subfigure
	
	\subfigure
	\centering \includegraphics[width=0.24\textwidth]{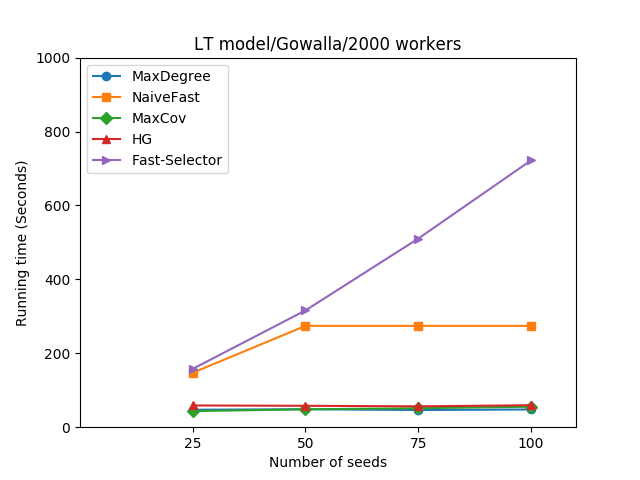}
	\includegraphics[width=0.24\textwidth]{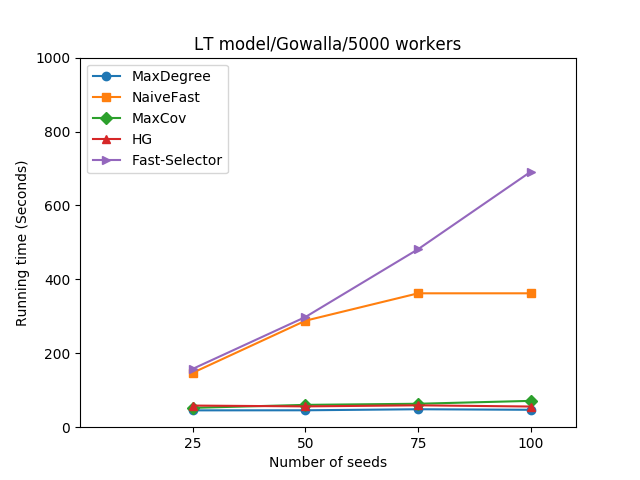}
	\subfigure
	
	\caption{Running time comparison for different models and datasets (under various settings in number of seeds and workers \label{fig_7}}
\end{figure}

Fig. ~\ref{fig_7} reports the running time of different algorithms. All results are measured on the reasonably efficient implementation of the various algorithms. Although \textit{Fast-Selector} needs longer running time than other heuristic algorithms, its running time is less than 1000 seconds on both two datasets. As the algorithm is executed offline, the computation time is acceptable. 

Moreover, we also verify the rationality of functions for two MCS-specific factors, where we vary the function $Cos()$ and $F()$ defined in Section 3.2. For topic similarity, we compare $Cos()$ with the Jaccard similarity function $J()$. For incentive attraction, we compare $F()$ with the linear function. We re-execute \textit{Fast-Selector} with different function combinations, and Table ~\ref{table_3} shows the coverage comparison of different combinations on two datasets. It indicates that the function choice is reasonable on these datasets. 

\begin{table}[!t]
	\renewcommand{\arraystretch}{1.3}
	\caption{Coverage comparison of different functions (For each dataset, it is the average result on different parameter settings.)}
	\label{table_3}
	\centering
	\begin{tabular}{c||c||c||c||c}
		\hline
		\bfseries & $Cos()$, $F()$ & $J()$, $F()$ & $Cos()$, Linear & $J()$, Linear \\
		\hline\hline
		Brightkite & 0.729 & 0.727 & 0.710 & 0.698 \\
		\hline
		Gowalla & 0.701 & 0.700 & 0.690 & 0.688 \\
		\hline
	\end{tabular}
\end{table}

\subsection{Basic-Selector vs Fast-Selector: a Comparison}
\begin{figure}
	\centering \includegraphics[width=1.7in]{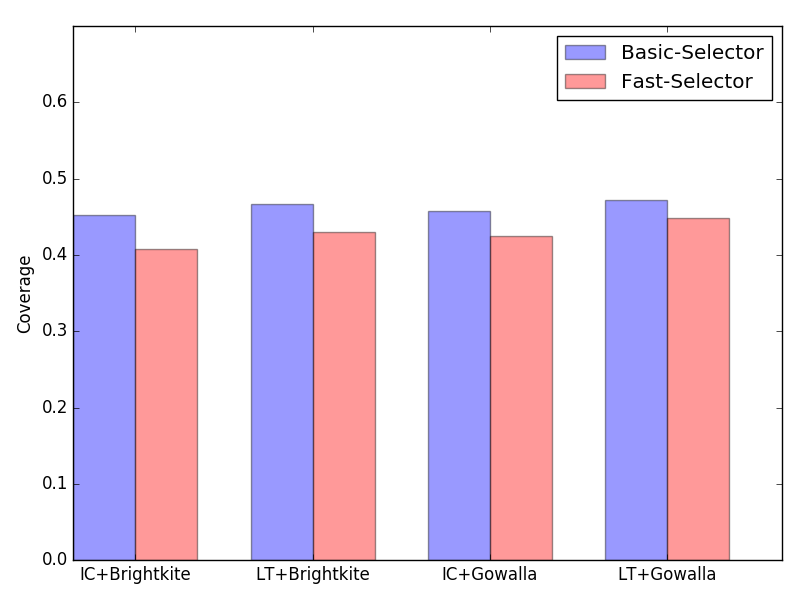}
	\includegraphics[width=1.7in]{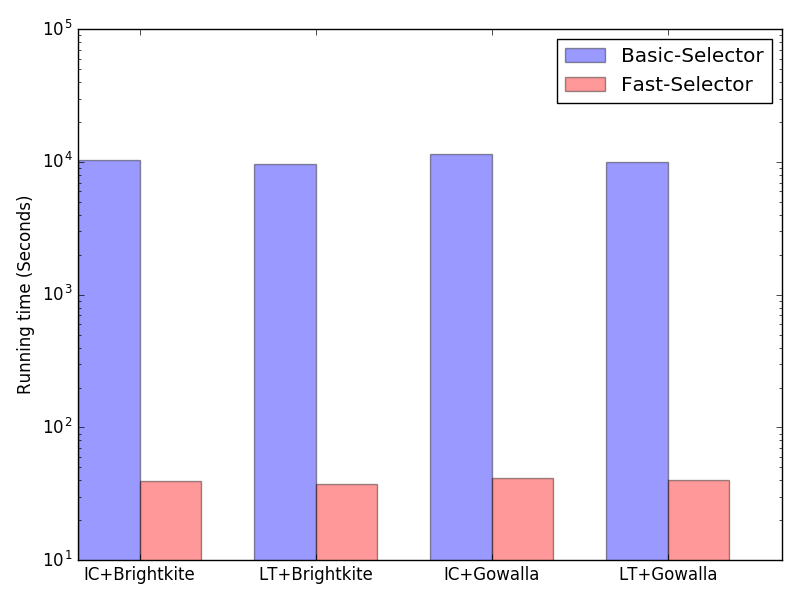}
	\caption{Coverage and running time comparison: \textit{Basic-Selector} vs \textit{Fast-Selector} (left: coverage; right: running time) \label{fig_8}}
\end{figure}

Fig. ~\ref{fig_8} (left) shows the coverage comparison between \textit{Basic-Selector} and \textit{Fast-Selector}, in which 1000 nodes are randomly selected from both two datasets. The results show that the \textit{Fast-Selector} achieves an average 6\%\textasciitilde9\% lower coverage compared to the \textit{Basic-Selector} in different combinations of datasets (Brightkite and Gowalla) and influence propagation models (extended IC and LT model). Meanwhile, Fig. ~\ref{fig_8} (right) shows the running time comparison, where it is observed that \textit{Basic-Selector} runs 100x slower than \textit{Fast-Selector}. Thus, we can see that \textit{Fast-Selector} is much more efficient than \textit{Basic-Selector} while only sacrificing a slight fraction of the coverage. 

The reason why the \textit{Fast-Selector} is more efficient is that it avoids most of the Monte-Carlo simulations adopted in \textit{Basic-Selector} when iteratively selecting seeds. Although the invoking of  $networkSpread(U^{'},q)$ in \textit{Fast-Selector} also brings extra computation cost, it requires much less computing time (10 simulations adopted by \textit{Fast-Selector}) compared to the Monte-Carlo simulation (typically 1000 simulations) used in \textit{Basic-Selector}. This is because $networkSpread(U^{'},q)$   only need to roughly judge the phase of the propagation (i.e., budget-sensitive or budget-insensitive), while the Monte-Carlo simulation needs to accurately predict the probability of each worker to be influenced.

\subsection{Impact of Budget Splitting for Seeds and Non-Seeds}
\begin{figure}
	\centering \includegraphics[width=3in]{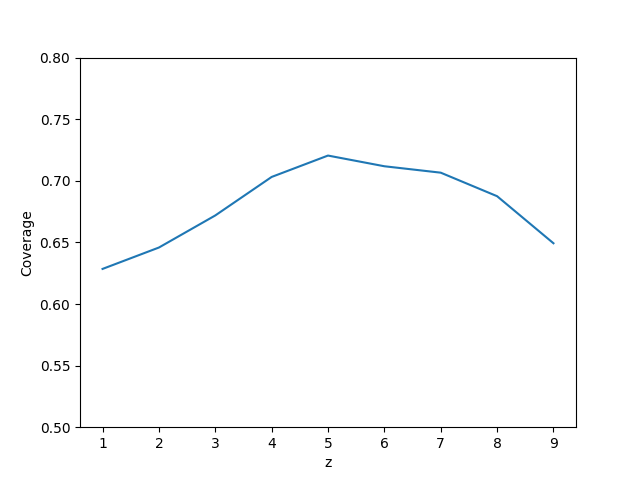}
	\caption{Impact of Budget Splitting for Seeds and Non-Seeds \label{fig_9}}
\end{figure}

In this paper, we assume that the incentive reward for each seed and non-seed worker is equal. However, in realistic scenarios, we may set different incentive rewards for them. Thus, we design extra experiments to see the performance when splitting the whole budget into seed rewards and non-seed rewards. We set the total budget as 600. The reward of each seed and non-seed worker is set as 2.0 and 1.0, respectively. The total budget for seed and non-seed workers is defined as $B_{sd}$ and $B_{nsd}$, respectively, which satisfies $B_{nsd}/B_{sd}=z$ and $B_{sd}+B_{nsd}=600.$ In this experiment, we vary the splitting of total budget (i.e., parameter $z$) for seed and non-seed users. The experimental results in Fig ~\ref{fig_9} show that different splitting of the total budget for seed and non-seed worker do have an impact on the finally achieved temporal-spatial coverage. It implies that careful analysis about the splitting should be considered to achieve optimal coverage in social-network-assisted MCS, which can be added to the future work.

\subsection{Summary and Implications}
By summarizing the above experimental results, we can draw the following conclusions. First,  \textit{Fast-Selector} obtains significantly higher temporal-spatial coverage than the baseline methods on both two datasets, for both two extended propagation models, and under various parameters settings. Second, \textit{Fast-Selector} is much more efficient than \textit{Basic-Selector}. Meanwhile, coverage achieved by \textit{Fast-Selector} is only slightly lower than \textit{Basic-Selector}.

In addition, we get the following implications for MCS and other location-relevant crowdsourcing systems.

First, \textit{in terms of MCS}, the social-network-assisted approach provides an alternative way for worker recruitment. In particular, in order to maximize the coverage, both the number of collaborative workers and their trajectory diversity are considered. Besides, the proposed algorithms should be appropriately adopted according to specific worker recruitment requirements. The advantage of \textit{Fast-Selector} is more obvious in large-scale social networks. Thus, when workers are recruited on a large-scale social network, \textit{Fast-Selector} is preferred (e.g. worker recruitment of an environmental monitoring task on Facebook, whose sensing area is a big city). However, if the size of the social network is not too large or the running time is not that crucial (e.g., worker recruitment of crowd detection task on a local community BBS, whose sensing area is a university campus),  \textit{Basic-Selector} is preferred as it can obtain a higher coverage. 

Second, \textit{ in terms of crowdsourcing systems}, our approach may inspire other types of location-based crowd work. Jumping out of the specific problem defined in this paper, one important inspiration of our work is the basic idea of jointly optimizing the number of crowd workers and their spatial diversity when recruiting them on the social networks. For instance, we spread a survey to people in a certain city, with the objective of jointly maximizing the volume and spatial diversity of returned answers. In this scenario, though the detailed algorithm should be re-designed, the basic insights in our work can be borrowed.
\section{Discussion}

This subsection discusses the limitations in this work, which can be added to our future work.

\subsection{More Sophisticated Task Acceptance Prediction Model}
In this paper, one fundamental issue is to predict the set of users who will accept the MCS task.  The inspiration of our work in this part lies in the insight that we should combine typical influence maximization models in social network research community with some MCS-specific factors. However, our proposed model has the following limitations. First, in addition to interest and incentives, some other factors (e.g., energy consumption and privacy concern) may also have an impact on the user's acceptance willingness. Second, the modeling of MCS-specific factors can be further explored. For example, cosine similarity may not be the best metric to measure the degree of interest matching. Third, although we propose an extended influence propagation model, it may not be the best if a real-world large-scale evaluation is executed. In summary, there are still two open research questions that need to be addressed in our future work: (a) what are the full set of factors influencing the MCS task propagation and acceptance. (b) How these factors are fused to build a more sophisticated predictive model.

\subsection{Improving the Seed Selection Algorithm}

Although the experiment indicates the effectiveness of our proposed algorithms, we should further improve them. For example, to avoid the time-consuming Monte Carlo simulation, \textit{Fast-Selector} uses a heuristic utility function for seed selection, where both the influence and geographical distribution of seeds are considered. Specifically, we use the degree of a user on the social network graph, a widely-used influence indicator \cite{24,25}, to measure users' influence. In the future work, we attempt to investigate and exploit  more sophisticated influence metrics in the future work. In addition, improving the location prediction  accuracy can further enhance the seeds selection. Our approach predicts the workers' location merely based on their own historical records. In fact, social relationships can be used to improve the prediction accuracy \cite{33}, which will be added to our future work. Another part that our work could improve upon in the future is to provide performance bounds on the basic and the fast algorithm.

\subsection{More Sophisticated Incentive Models}
To simplify the problem, we assume that each recruited worker gets an equal incentive reward. As a matter of fact, we could explore more complicated incentive models in the future work. For example, we could assign higher reward for seed nodes than non-seed node. We can also separate the task completion from task propagation. Currently, we assume that each recruited worker will accept and complete the task, and meanwhile recommending it to their friends. Actually, we can make it more flexible. For example, if a user does not have time to complete a task but can recommend it to his/her friends, he/she can also earn a certain proportion of the rewards. In summary, the incentive models can be more sophisticated, which should be designed according to different types of MCS tasks and social network apps.

\subsection{Social Network Data Availability}
When implementing the proposed idea on real-world systems, there are challenges in terms of data availability. (1) \textit{Cooperative Scenarios}. That is, the service providers of the mobile social network are the collaborators of the MCS worker recruitment for common interest (e.g., for social benefit, or advertisement fee, etc.), and they will actively provide the needed social metadata (e.g., friend relationships, profiles, check-in locations, etc.) after certain privacy-preserving operations (e.g., anonymization). (2) \textit{Non-cooperative Scenarios}. In this case, the service providers are not the collaborators, so that we need to extract the metadata automatically from online social networks, which is much more complicated. Many social network apps provide open APIs for third-party developers, through which we can extract various types of users’ information needed (e.g., friends list, preference list, relative distance). However, even with these open APIs, it is still challenging to efficiently extract information while keeping the legality in terms of privacy preserving policy adopted by service providers.

\subsection{Evaluation and Refinement with Large-Scale User Study}
Although we use real-world social network datasets (including real friendship and trajectories) to evaluate our approach, the major limitation, which are also suffered in the state-of-the-art research work in influence maximization \cite{21,22,24,25,26}, lie in the fact that a number of parameters of the task attributes and propagation process are simulated with certain assumptions. For example, we collect additional data from 93 online volunteers to and produce distributions of parameters, and then generate simulated workers according to these distributions and treat the generated workers as the subjects in the two open datasets (i.e., Brightkite and Gowalla). However, this process is imperfect because the generated workers may not necessarily represent the original subjects in two datasets. Therefore, we are now collaborating with a local social network (i.e., a city-scale online forum with social network structures) to evaluate our solution in real-world settings. Specifically, we attempt to publish some MCS tasks on this forum and introduce our algorithms to recruit workers. We believe such evaluations can help identify more practical issues, which will iteratively refine our proposed models and algorithms.

\section{Conclusion}
In this paper, we propose an MCS task worker recruitment solution based on the mobile social network. Our approach first selects a subset of users on the social network as initial seeds and pushes the task to them. Then influenced users would accept and propagate it to friends. Specifically, two seeds selection algorithms are proposed, and experiments on real-world social datasets shows the effectiveness. Finally, we point out the limitations in models and evaluations with future work directions


\begin{IEEEbiography}[{\includegraphics[width=1in,height=1.25in,clip,keepaspectratio]{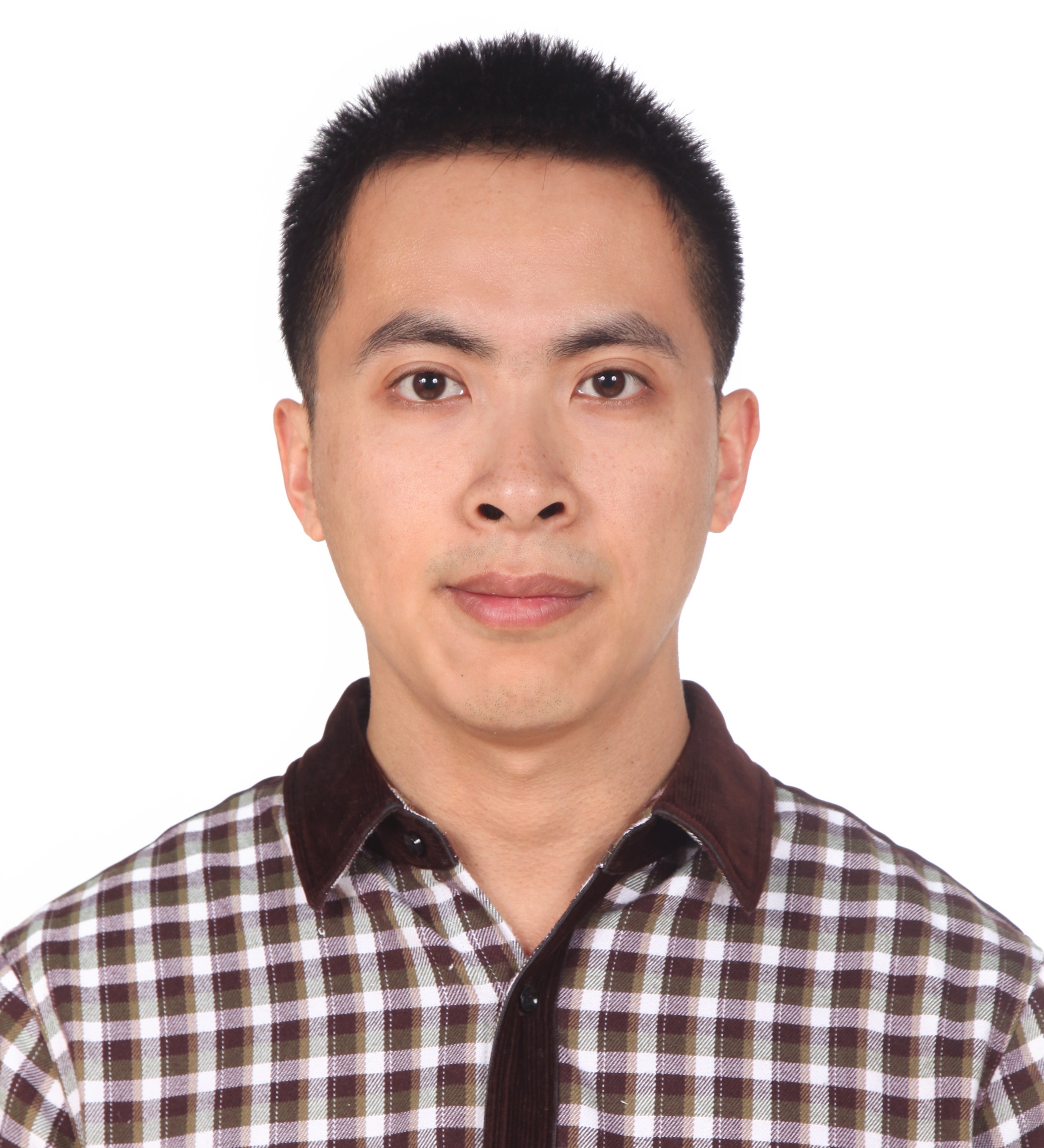}}]{Jiangtao Wang}

	Jiangtao Wang received his Ph.D. degree in Peking University, Beijing, China, in 2015. He is currently an assistant professor in Institute of Software, School of Electronics Engineering and Computer Science, Peking University. His research interest includes collaborative sensing, mobile computing, and ubiquitous computing.

\end{IEEEbiography}

\begin{IEEEbiography}[{\includegraphics[width=1in,height=1.25in,clip,keepaspectratio]{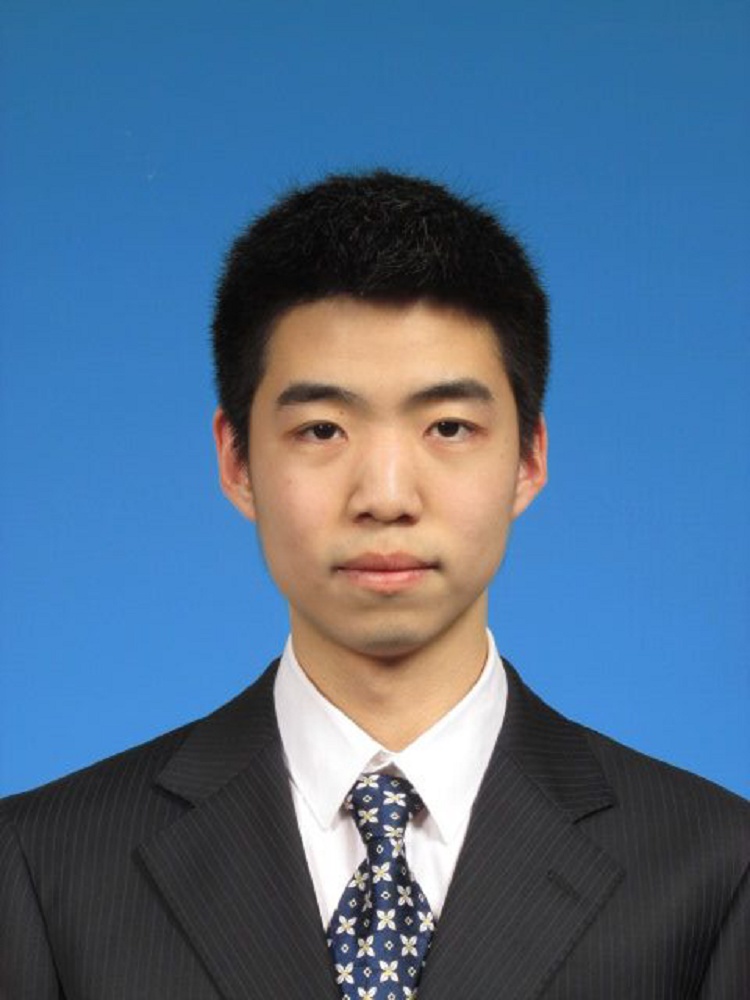}}]{Feng Wang}

	Feng Wang is a master student at School of Electronic Engineering and Computer Science, Peking University, China. His research interest is mobile crowd sensing.

\end{IEEEbiography}

\begin{IEEEbiography}[{\includegraphics[width=1in,height=1.25in,clip,keepaspectratio]{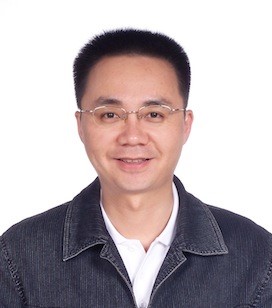}}]{Yasha Wang}

	Yasha Wang received his Ph.D. degree in Northeastern University, Shenyang, China, in 2003. He is a professor and associate director of National Research \& Engineering Center of Software Engineering in Peking University, China. His research interest includes urban data analytics, ubiquitous computing, software reuse, and online software development environment. He has published more than 50 papers in prestigious conferences and journals, such as ICWS, UbiComp, ICSP and etc. As a technical leader and manager, he has accomplished several key national projects on software engineering and smart cities. Cooperating with major smart-city solution providing companies, his research work has been adopted in more than 20 cities in China.

\end{IEEEbiography}

\begin{IEEEbiography}[{\includegraphics[width=1in,height=1.25in,clip,keepaspectratio]{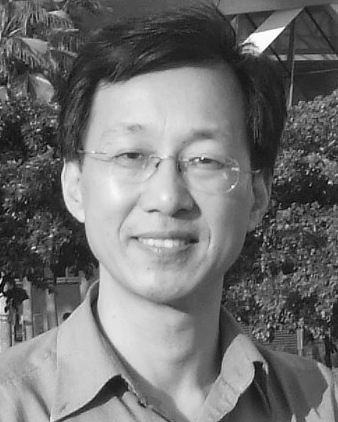}}]{Daqing Zhang}

	Daqing Zhang is a professor at Peking University, China, and T\'{e}l\'{e}com SudParis, France. He obtained his Ph.D from the University of Rome “La Sapienza,” Italy, in 1996. His research interests include context-aware computing, urban computing, mobile computing, and so on. He served as the General or Program Chair for more than 10 international conferences. He is an Associate Editor for ACM Transactions on Intelligent Systems and Technology, IEEE Transactions on Big Data, and others.

\end{IEEEbiography}

\begin{IEEEbiography}[{\includegraphics[width=1in,height=1.25in,clip,keepaspectratio]{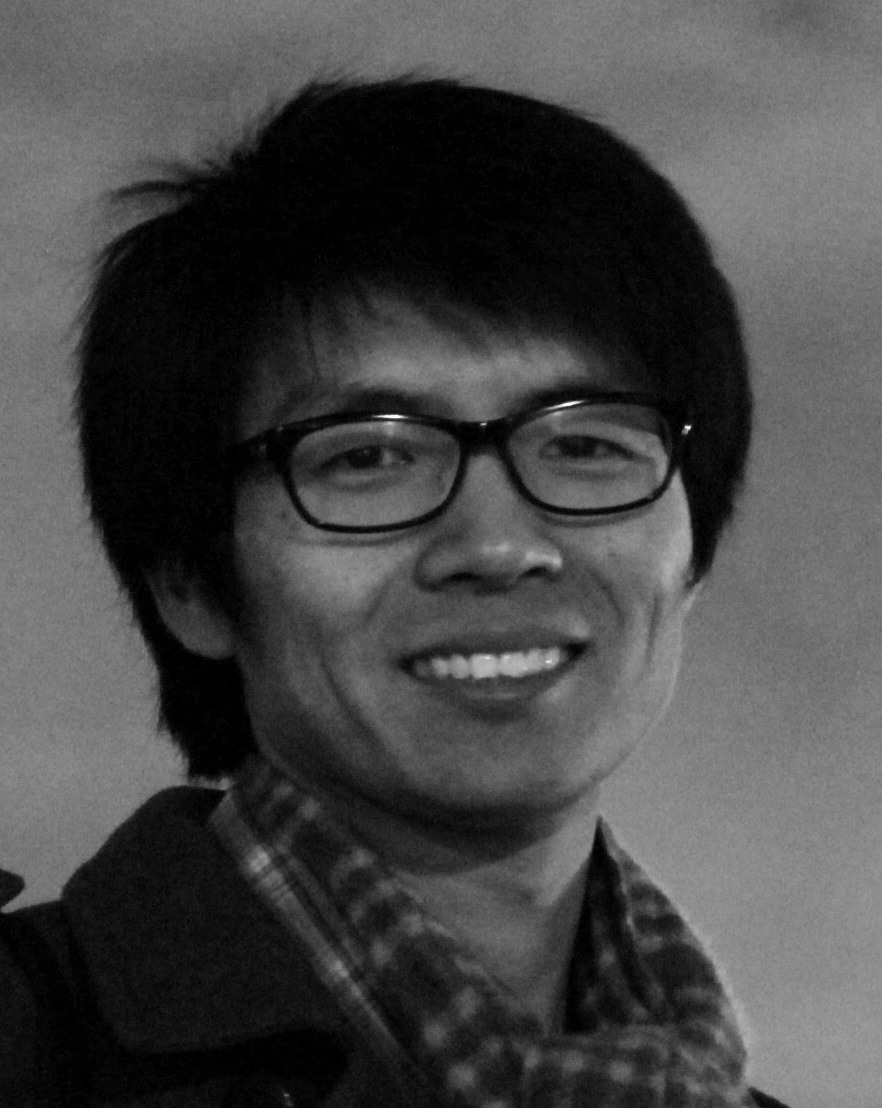}}]{Leye Wang}

	LEYE WANG obtained his Ph.D. from Institut Mines-T\'{e}l\'{e}com/T\'{e}l\'{e}com SudParis and Universit\'{e} Pierre et Marie Curie, France, in 2016. He received his M.Sc. and B.Sc. in computer science from Peking university, China. His research interests include mobile crowdsensing, social networks, and intelligent transportation systems.

\end{IEEEbiography}

\begin{IEEEbiography}[{\includegraphics[width=1in,height=1.25in,clip,keepaspectratio]{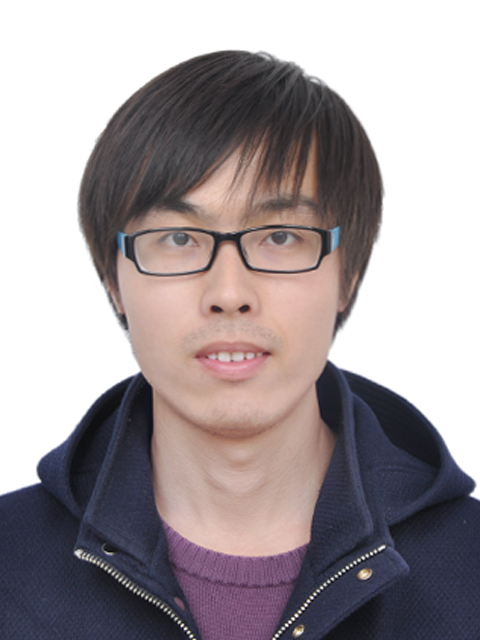}}]{Zhaopeng Qiu}

	Zhaopeng Qiu is an undergraduate student at School of Electronic Engineering and Computer Science, Peking University, China. His research interest is data analysis and NLP.

\end{IEEEbiography}


%\vfill



\begin{thebibliography}{1}
	
	\bibitem{1}
	Howe, J. (2006). The rise of crowdsourcing. Wired Magazine, 14(6), 1-4.
	
	\bibitem{2}
	Kittur, A., Nickerson, J. V., Bernstein, M., Gerber, E., Shaw, A., Zimmerman, J., ... \& Horton, J. (2013, February). The future of crowd work. In Proceedings of the 2013 conference on Computer supported cooperative work, pp. 1301-1318. ACM.
	\bibitem{3}
	Ganti, R. K., Ye, F., \& Lei, H. (2011). Mobile crowdsensing: current state and future challenges. IEEE Communications Magazine, 49(11), 32-39.
	
	\bibitem{4}
	Wang, J., Wang, Y., Zhang, D., Wang, F., He, Y., \& Ma, L. (2017, February). PSAllocator: multi-task allocation for participatory sensing with sensing capability constraints. In \textit{Proceedings of the 2017 ACM Conference on Computer Supported Cooperative Work and Social Computing} (pp. 1139-1151). ACM.
	
	\bibitem{5}
	S. Reddy, D. Estrin, and M. Srivastava. Recruitment framework for participatory sensing data collections. In Proceedings of Pervasive, pages 138-155. 2010.
	
	\bibitem{6}
	Adish Singla and Andreas Krause. Incentives for privacy tradeoff in community sensing. In First AAAI Conference on Human Computation and Crowdsourcing, 2013.
	
	\bibitem{7}
	Giuseppe Cardone, Luca Foschini, Paolo Bellavista, Antonio Corradi, Cristian Borcea, Manoop Talasila, and Reza Curtmola. Fostering participaction in smart cities: a geo-social crowdsensing platform. Communications Magazine, IEEE, 51(6), 2013.
	
	\bibitem{8}
	H. Xiong, D. Zhang, L. Wang, J. Gibson, and J. Zhu, "EEMC: Enabling energy-efficient mobile crowdsensing with anonymous workers," ACM Transactions on Intelligent Systems and Technology (TIST), 2015
	
	\bibitem{9}
	Yadav, A., Wilder, B., Rice, E., Petering, R., Craddock, J., Yoshioka-Maxwell, A., ... \& Woo, D. (2017, May). Influence maximization in the field: The arduous journey from emerging to deployed application. In Proceedings of the 16th Conference on Autonomous Agents and Multi-Agent Systems (pp. 150-158). International Foundation for Autonomous Agents and Multi-agent Systems.
	
	\bibitem{10}
	Karaliopoulos, M., Telelis, O., \& Koutsopoulos, I. (2015). User Recruitment for Mobile Crowdsensing over Opportunistic Networks. In INFOCOM 2015 Proceedings, IEEE
	
	\bibitem{11}
	S. Hachem, A. Pathak, and V. Issarny. Probabilistic registration for large-scale mobile participatory sensing. In Proceedings of the 2013 IEEE International conference on Pervasive Computing and Communications, volume 18, page 22, 2013..
	
	\bibitem{12}
	Zhang, D., Wang, L., Xiong, H., \& Guo, B. (2014). 4W1H in mobile crowd sensing. IEEE Communications Magazine, 52(8), 42-48.
	
	\bibitem{13}
	Yan Liu, Bin Guo, Yang Wang, Wenle Wu, Zhiwen Yu, and Daqing Zhang. 2016.TaskMe: multi-task allocation in mobile crowd sensing. In Proceedings of the 2016ACM International Joint Conference on Pervasive and Ubiquitous Computing (UbiComp'16).
	
	\bibitem{14}
	T. Kandappu, N. Jaiman, R. Tandriansyah, A. Misra, S. F. Cheng, C. Chen, H. C.Lau, D. Chander, K. Dasgupta, "TASKer: Behavioral Insights via Campus-based Experimental Mobile Crowdsourcing", 2016 ACM International Joint Conference on Pervasive and Ubiquitou Computing (UbiComp 2016), September 2016
	
	\bibitem{15}Cho, E., Myers, S. A., \& Leskovec, J. (2011, August). Friendship and mobility: user movement in location-based social networks. In Proceedings of the 17th ACM SIGKDD international conference on Knowledge discovery and data mining (pp. 1082-1090). ACM.
	
	\bibitem{16}
	Case, D. O. (2012). Looking for information: A survey of research on information seeking, needs and behavior. Emerald Group Publishing.
	
	\bibitem{17}
	H. Xiong, D. Zhang, G. Chen, L. Wang, and V. Gauthier, "Crowdtasker: Maximizing coverage quality in piggyback crowdsensing under budget constraint," in IEEE International Conference on Pervasive Computing and Communications (Percom'15), 2015.
	
	\bibitem{18}
	Zhang, M., Yang, P., Tian, C., \& Tang, S. (2015). Quality-aware sensing coverage in budget constrained mobile crowdsensing networks. IEEE Transactions on Vehicular Technology, 1-1.
	
	\bibitem{19}
	D. Zhang, H. Xiong, L. Wang, and G. Chen, "Crowdrecruiter: selecting workers for piggyback crowdsensing under probabilistic coverage constraint", in The 2014 ACM Conference on Ubiquitous Computing, UbiComp'14, Seattle, WA, USA, 2014, pp. 703-714.
	
	\bibitem{20}
	H. Xiong, D. Zhang, L. Wang, and H. Chaouchi, "EMC3: Energy-efficient data transfer in mobile crowdsensing under full coverage constraint," IEEE Transactions on Mobile Computing, 2015.
	
	\bibitem{21}
	D. Kempe, J. Kleinberg, and E. Tardos, "Maximizing the spread of influence through a social network," in Proc. 9th ACM SIGKDD Int. Conf. Knowl. Discovery Data Mining, 2003, pp. 137-146.
	
	\bibitem{22}
	D. Kempe, J. Kleinberg, and E. Tardos, "Influential nodes in a diffusion model for social networks," in Proc. 32nd Int. Colloq. Automata, Language Programm., 2005, no. 32, pp. 1127-1138.
	
	\bibitem{23}
	Mooney, C. Z. (1997). \textit{Monte carlo simulation} (Vol. 116). Sage Publications.
	
	\bibitem{24}
	W. Chen, Y. Wang, and S. Yang, "Efficient influence maximization in social networks," in Proc. 15th ACM SIGKDD Int. Conf. Knowl. Discovery Data Mining, 2009, pp. 199-208.
	
	\bibitem{25}
	Wang, Y., Cong, G., Song, G., \& Xie, K. (2010, July). Community-based greedy algorithm for mining top-k influential nodes in mobile social networks. In Proceedings of the 16th ACM SIGKDD international conference on Knowledge discovery and data mining (pp. 1039-1048). ACM. 
	
	\bibitem{26}
	Song, G., Zhou, X., Wang, Y., \& Xie, K. (2015). Influence maximization on large-scale mobile social network: a divide-and-conquer method. IEEE Transactions on Parallel and Distributed Systems, 26(5), 1379-1392.
	
	\bibitem{27}Kaiyu Feng, Gao Cong, Sourav S. Bhowmick, Shuai Ma. In Search of Influential Event Organizers in Online Social Networks. In SIGMOD 2014, Pages 63-74.
	
	\bibitem{28}Wei Lu, Wei Chen, Laks V. S. Lakshmanan: From Competition to Complementarity: Comparative Influence Diffusion and Maximization. PVLDB 9(2): 60-71, 2015.
	
	\bibitem{29}
	Cho, Y. S., Ver Steeg, G., \& Galstyan, A. (2014, June). Where and Why Users "Check In". In AAAI 2014 (pp. 269-275). 
	
	\bibitem{30}
	Yang, D., Xue, G., Fang, G., \& Tang, J. (2015). Incentive mechanisms for crowdsensing: crowdsourcing with smartphones. IEEE/ACM Transactions on Networking, 1-13.
	
	\bibitem{31}
	Zhipeng Cai, Mingyuan Yan, Yingshu Li. Using crowdsourced data in location-based social networks to explore influence maximization. IEEE INFOCOM 2016.
	
	\bibitem{32}
	Lane, N. D., Chon, Y., Zhou, L., Zhang, Y., Li, F., \& Kim, D., et al. (2013). Piggyback CrowdSensing (PCS): energy efficient crowdsourcing of mobile sensor data by exploiting smartphone app opportunities. ACM Conference on Embedded Networked Sensor Systems (pp.1-14).
	
	\bibitem{33}
	Sadilek, A., Kautz, H., \& Bigham, J. P. (2012, February). Finding your friends and following them to where you are. In Proceedings of the fifth ACM international conference on Web search and data mining (pp. 723-732). ACM.
	
	\bibitem{34}
	https://sojump.com/jq/15291868.aspx (survey, English version); https://sojump.com/jq/14950858.aspx (survey, Chinese  version)

	\bibitem{35}
	Cardone, G., Cirri, A., Corradi, A., \& Foschini, L. (2014). The participact mobile crowd sensing living lab: The testbed for smart cities. IEEE Communications Magazine, 52(10), 78-85.

	\bibitem{36}
	Bellavista, P., Corradi, A., Foschini, L., \& Ianniello, R. (2015). Scalable and cost-effective assignment of mobile crowdsensing tasks based on profiling trends and prediction: The participact living lab experience. Sensors, 15(8), 18613-18640.

	\bibitem{37}
	Cardone, G., Corradi, A., Foschini, L., \& Ianniello, R. (2016). Participact: A large-scale crowdsensing platform. IEEE Transactions on Emerging Topics in Computing, 4(1), 21-32.
	
	\bibitem{38}
	Chessa, S., Girolami, M., Foschini, L., Ianniello, R., Corradi, A., \& Bellavista, P. (2017). Mobile crowd sensing management with the ParticipAct living lab. Pervasive and Mobile Computing, 38, 200-214.

	\bibitem{39}
	1.	Dehaene, S. (2003). The neural basis of the Weber–Fechner law: a logarithmic mental number line. Trends in cognitive sciences, 7(4), 145-147.

	\bibitem{40}
	Zhao, Z., Cheng, J., Wei, F., Zhou, M., Ng, W., \& Wu, Y. (2014, November). Socialtransfer: Transferring social knowledge for cold-start cowdsourcing. In Proceedings of the 23rd ACM International Conference on Conference on Information and Knowledge Management (pp. 779-788). ACM.

\end{thebibliography}
\end{document}